\documentclass[prd,preprint,tightenlines,superscriptaddress,floatfix,showpacs,preprintnumbers,nofootinbib,eqsecnum]{revtex4}

 \usepackage[dvips,final]{graphicx}
     \usepackage{epsfig}
      \usepackage{bm}

\def\Pom{{\bf I\!P}}
\def\Reg{{\bf I\!R}}
\newcommand{\p}{\partial}

\newcommand{\twosidep}[1]{\stackrel{\leftrightarrow}{\p^{#1}}}

\begin{document}

\thispagestyle{empty} \preprint{\hbox{}} \vspace*{-10mm}

\title{Subleading processes in production\\ 
of $W^+ W^-$ pairs in proton-proton collisions}

\author{M.~{\L}uszczak}
\email{luszczak@univ.rzeszow.pl}
\affiliation{University of Rzesz\'ow, PL-35-959 Rzesz\'ow, Poland}

\author{A.~Szczurek}
\email{antoni.szczurek@ifj.edu.pl}

\affiliation{Institute of Nuclear Physics PAN, PL-31-342 Cracow, Poland} 
\affiliation{University of Rzesz\'ow, PL-35-959 Rzesz\'ow, Poland}

\date{\today}

\begin{abstract}
We discuss many new subleading processes for inclusive production of 
$W^+ W^-$ pairs not included in the literature so far. 
We focus on photon-photon induced processes.
We include elastic-elastic, elastic-inelastic, inelastic-elastic 
and inelastic-inelastic contributions. 
The inelastic photon distributions in the proton are 
calculated in two different ways: naive approach used already in 
the literature and using photon distributions by solving special 
evolution equation with photon being a parton in the proton. 
The results strongly depend on the approach used. 
We calculate also contributions with resolved photons. 
The diffractive components have similar characteristics as 
the photon-photon elastic-inelastic and inelastic-elastic mechanisms.
The subleading contributions are compared with the well known 
$q \bar q$ and $g g$ as well as with double-parton scattering contributions.
Predictions for the total cross section and differential
distributions in $W$- boson rapidity and transverse momentum as well
as $WW$ invariant mass are presented. The $\gamma \gamma$ components
constitute only about 1-2 \% of the inclusive $W^+ W^-$ cross section
but about 10 \% at large $W^{\pm}$ transverse momenta, and are even
comparable to the dominant $q \bar q$ component at large $M_{WW}$,
i.e. are much larger than the often celebrated $g g \to W^+ W^-$ component.
Its size is comparable to double parton scattering contribution.
Only elastic-elastic, elastic-inelastic and inelastic-elastic contributions 
could be potentially measured to verify our predictions using forward
proton detectors.
\end{abstract}

\maketitle

\section{Introduction}

The reaction $p p \to W^+ W^- X$ constitutes important, irreducible
background to the observation of the Higgs boson in the $W^+ W^-$
channel. Furthermore
it can be used to test Standard Model gauge boson couplings
and study them in models beyond Standard Model.

The $\gamma \gamma \to W^+ W^-$ process is interesting 
reaction to test the Standard Model and any other theories beyond 
the Standard Model.
The photon-photon contribution for the purely exclusive production 
of $W^+ W^-$ was considered recently in the literature 
\cite{royon,piotrzkowski}.
The exclusive diffractive mechanism of central exclusive production
of $W^+W^-$ pairs in proton-proton collisions at the LHC 
(in which diagrams with intermediate virtual Higgs boson as well as quark box
diagrams are included) was discussed in Ref.~\cite{LS2012} and turned 
out to be negligibly small.
The diffractive production and decay of Higgs boson into the $W^+W^-$ pair 
was also discussed in Ref.~\cite{WWKhoze}. 
Provided this is the case, the $W^+W^-$ pair production signal would
be particularly sensitive to New Physics contributions in 
the $\gamma \gamma \to W^+ W^-$ subprocess \cite{royon,piotrzkowski}. 
Similar analysis has been considered recently
for $\gamma \gamma \to Z Z$ \cite{Gupta:2011be}. 
Corresponding measurements will be possible to perform at the ATLAS or CMS
detectors with the use of very forward proton detectors 
\cite{forward_protons}. 

In the present paper we concentrate on inclusive production of $W^+ W^-$ pairs.
The production of $W^+ W^-$ has been measured recently with 
the CMS and ATLAS detectors \cite{CMS2011, ATLAS2012}.
The total measured cross section with the help of the CMS detector is  
41.1 $\pm$ 15.3 (stat) $\pm$ 5.8 (syst) $\pm$ 4.5 (lumi) pb, 
the total measured cross section with the ATLAS detector with slightly
better statistics is
54.4 $\pm$  4.0 (stat.) $\pm$  3.9 (syst.) $\pm$  2.0 (lumi.) pb. 
The more precise ATLAS result is somewhat bigger than the Standard Model
predictions of 44.4  $\pm$  2.8 pb \cite{ATLAS2012}.

The double parton scattering (DPS) mechanism of $W^+ W^-$ production
was discussed e.g. in 
Ref.\cite{KS2000,Kulesza2010,GKKS2011}. 
The $W^+ W^-$ final states constitutes a background to Higgs production.
It was discussed recently that double-parton scattering could explain
a large part of the observed signal \cite{KP2013}. We shall also discuss
the double parton scattering mechanism in the present paper.

It is our aim here to focus on the role of photon-photon induced processes.
Very recently the CMS and D0 collaborations measured semi-exclusive production
of $W^+ W^-$ pairs \cite{gamgam_WW_CMS, gamgam_WW_D0}. 
From the theoretical side so far only purely exclusive photon-photon reaction 
was studied in this context in the literature \cite{royon,piotrzkowski,LS2012}. 
In the present paper we wish to include also photon-induced inelastic 
processes in which photon emission breaks at least one of the photons.
This will be performed in two different methods.
Only the inelastic-inelastic contribution was discussed very recently
in a broader context of electroweak corrections \cite{electroweak}. 

Furthermore we shall include for the first time processes with resolved 
photons as well as single-diffractive production of $W^+ W^-$ pairs.
The two mechanisms have rather similar characteristics of the final state
and could be studied experimentally separately with the help of 
forward proton detectors.

In principle, also production of the Higgs boson with its decay into
$W W^*$ (one real, one virtual) channel may contribute. 
The present experimental results on Higgs production 
\cite{Higgs_ATLAS,Higgs_CMS} give its mass of about $M_H$ = 125 GeV and
strongly suggest that the observed Higgs boson is almost consistent 
with the Standard Model.
Then the corresponding contribution is very small.
In our present approximation of two on-shell W bosons this contribution
vanishes. 

We shall calculate phase space integrated cross section and
distributions in $W$-boson rapidity, transverse momentum as well as
in $M_{WW}$ invariant mass.
For reference we calculate also corresponding total cross section
and differential distributions for $q \bar q \to W^+ W^-$
and $g g \to W^+ W^-$ initiated subprocesses.
Since we concentrate on the missing mechanisms the latter will be
calculated in the leading-order approximation only, dispite calculations
in the next-to-leading order have been performed by different groups
\cite{WW_NLO}.

\section{$\gamma \gamma \to W^+ W^-$ reaction}

Let us start from a reminder about the $\gamma \gamma \to W^+ W^-$
coupling within the Standard Model.
The three-boson $WW \gamma$ and four-boson $WW \gamma\gamma$ couplings,
which contribute to the $\gamma \gamma \to W^+ W^-$ process in
the leading order read
\begin{eqnarray}
{\cal L}_{WW\gamma} & = &
-ie( A_\mu  W^-_\nu \twosidep{\mu} W^{+\nu}
+   W_\mu^- W^+_\nu \twosidep{\mu} A^\nu
+    W^+_\mu  A_\nu \twosidep{\mu} W^{-\nu}) \, ,
\nonumber \\ 
{\cal L}_{WW\gamma\gamma} & = &
-e^2\left(  W^{-}_{\mu} W^{+\mu}A_{\nu} A^{\nu} 
     - W^{+}_{\mu}A^{\mu} W^{-}_{\nu} A^\nu \right) \, ,
\label{eq:anom:lagrww2}
\end{eqnarray}
where the asymmetric derivative has the form
$X\twosidep{\mu}Y=X\p^{\mu}Y-Y\p^{\mu}X$.

\begin{figure*}
\begin{center}
\includegraphics[width=5cm]{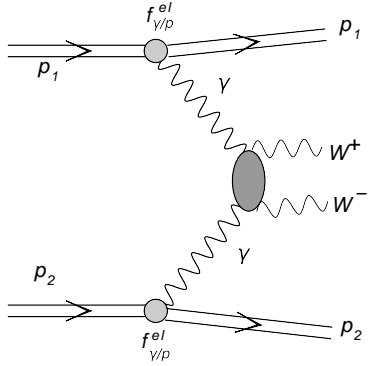}
\caption{The general diagram for the $pp \to pp W^+ W^-$ reaction
via $\gamma \gamma \to W^+ W^-$ subprocess.
}
\label{fig:m1}
\end{center}
\end{figure*}

\begin{figure*}
\begin{center}
\includegraphics[width=4cm]{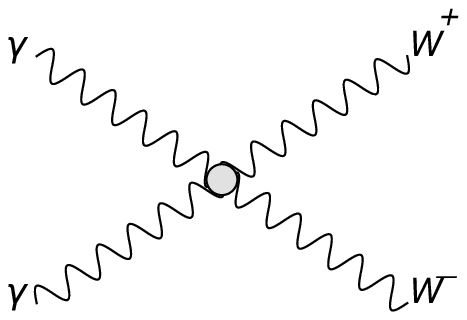}
\includegraphics[width=4cm]{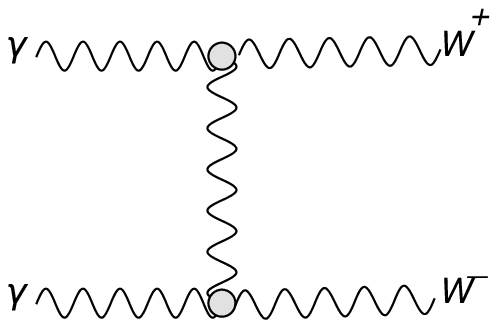}
\includegraphics[width=4cm]{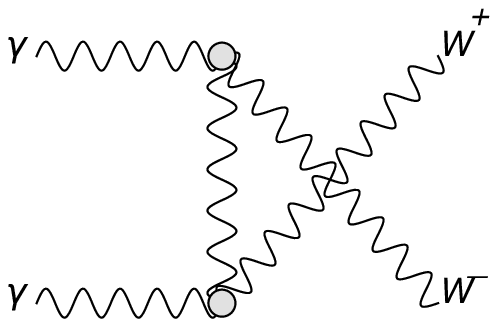}
\end{center}
\caption{The leading order $\gamma \gamma \to W^+ W^-$ subprocesses.
}
\label{fig:LO_subprocesses}
\end{figure*}

The general diagram for the exclusive reaction is shown in Fig.~\ref{fig:m1}.
The relevant leading-order subprocess diagrams are shown 
in Fig.~\ref{fig:LO_subprocesses}.

Then within the Standard Model, the elementary tree-level cross
section for the $\gamma \gamma \to W^+ W^-$ subprocess can be written 
in the very compact form in terms of the Mandelstam variables 
(see e.g. Ref.~\cite{DDS95})

\begin{equation}
\frac{d\hat{\sigma}}{d \Omega} = \frac{3 \alpha^2 \beta}{2\hat{s}} \left(
1 - \frac{2 \hat{s} (2\hat{s}+3m_W^2)}{3 (m_W^2 - \hat{t}) (m_W^2 -
\hat{u})} + \frac{2 \hat{s}^2(\hat{s}^2+ 3m_W^4)}{3 (m_W^2 -
\hat{t})^2(m_W^2 - \hat{u})^2} \right) \, ,
\label{gamgam_WW}
\end{equation}
where $\beta=\sqrt{1-4m_W^2/\hat{s}}$ is the velocity of the $W$
bosons in their center-of-mass frame and the electromagnetic
fine-structure constant $\alpha=e^{2}/(4\pi) \simeq 1/137$. 
The total elementary cross section can be obtained
by integration of the differential cross section above.

\section{Exclusive $p p \to p p W^+ W^-$ reaction}

The $p p \to p p W^+ W^-$ reaction is particularly interesting in the
context of $\gamma \gamma W W$ coupling \cite{royon,piotrzkowski}.

In the Weizs\"acker-Williams approximation,
the total cross section for the $pp \to pp (\gamma \gamma) \to W^+ W^-$
can be written as in the parton model:
\begin{equation}
\sigma = \int d x_1 d x_2 \, f_1^{WW}(x_1) \, f_2^{WW}(x_2) \,
\hat{\sigma}_{\gamma \gamma \to W^+ W^-}(\hat s) \, .
\label{EPA}
\end{equation}
We take the Weizs\"acker-Williams equivalent photon fluxes in
protons from Ref.~\cite{DZ}.

To calculate differential distributions the following parton formula
can be conveniently used
\begin{equation}
\frac{d\sigma}{d y_+ d y_- d^2 p_{W\perp}} = \frac{1}{16 \pi^2 {\hat s}^2}
\, x_1 f_1^{WW}(x_1) \, x_2 f_2^{WW}(x_2) \,
\overline{ | {\cal M}_{\gamma \gamma \to W^+ W^-}(\hat s, \hat t, \hat u)
  |^2} \, .
\label{EPA_differential}
\end{equation}
We shall not discuss here any approach beyond the Standard Model.
A potentially interesting Higgsless scenario of the
production of $W^+ W^-$ pairs has been discussed previously e.g. in
Refs.~\cite{royon,piotrzkowski}.

The exclusive cross section could be calculated also more precisely
in four-body calculation with corresponding $2 \to 4$ matrix element.
However, such a precision is not needed now when reviewing all potentially
important contributions. This may become important for large statistics 
analysis of the 14 TeV data with extra measurement of forward protons
and when discussing experimental cuts.

\section{Inclusive production of $W^+W^-$ pairs}
\label{sec:inclusive}

The dominant contribution of $W^+W^-$ pair production is initiated by
quark-antiquark annihilation \cite{DDS95}.
The gluon-gluon contribution to the inclusive cross section 
was calculated first in Ref.~\cite{gg_WW}.

Therefore for a comparison we also consider quark-antiquark and 
gluon-gluon components to the inclusive cross section. They will 
constitute a reference point for our calculations of 
the two-photon contributions.

\subsection{$q \bar q \to W^+ W^-$ mechanism}

\begin{figure*}
\begin{center}
\includegraphics[width=5cm]{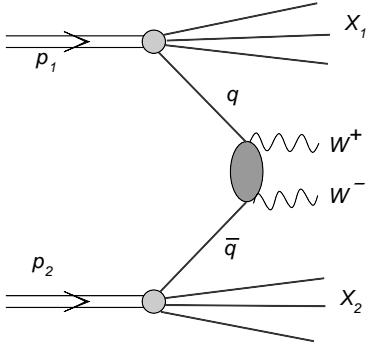}
\caption{A generic diagram representing mechanisms 
for production of $W^+W^-$ pairs in the
$q \bar q \to W^+ W^-$ subprocess.
}
\label{fig:qqbar_WW}
\end{center}
\end{figure*}

The generic diagram for the $q \bar q$ initiated processes
is shown in Fig.\ref{fig:qqbar_WW}. This contains $t$- and $u$-channel quark
exchanges as well as $s$-channel photon and $Z$-boson exchanges
\cite{DDS95}.

Therefore this process is also of interest as a probe of 
the gauge structure of the electroweak interactions.
Relevant leading-order matrix element, averaged over quark colors 
and over initial spin polarizations and summed over final spin 
polarization, can be found e.g. in Ref.~\cite{Eichten}.

The corresponding differential cross section in leading-order
approximation can be written as:
\begin{equation}
\frac{d\sigma}{d y_+ d y_- d^2 p_{W\perp}} = \frac{1}{16 \pi^2 {\hat s}^2}
\,[ x_1 q(x_1,\mu^2) \, x_2 {\bar q}(x_2,\mu^2) +
    x_1 {\bar q}(x_1,\mu^2) \, x_2 q(x_2,\mu^2) ]
\overline{ | {\cal M}_{q {\bar q} \to W^+ W^-}(\hat s, \hat t, \hat u)
  |^2} \, .
\label{qqbar_annihilation}
\end{equation}
%


\subsection{$g g \to W^+ W^-$ mechanism}

\begin{figure*}
\begin{center}
\includegraphics[width=5cm]{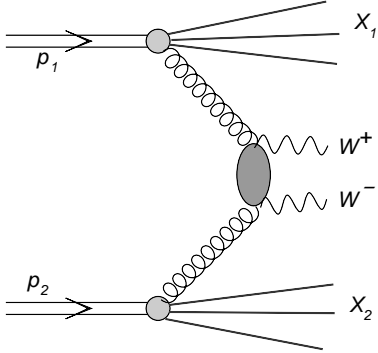}
\caption{A generic diagram representing mechanisms 
for production of $W^+W^-$ pairs in the
$gg \to W^+ W^-$ subprocess.
}
\label{fig:gg_WW}
\end{center}
\end{figure*}

The generic diagram for the $g g$ initiated processes 
is shown in Fig.\ref{fig:gg_WW}. This contains both quark box diagrams
and heavy-quark triangle with s-channel Higgs boson in the
intermediate stage. More details of the relevant calculation can be
found e.g. in Ref.\cite{LS2012}.

The corresponding differential cross section corresponding
to this contribution can be written as:
\begin{equation}
\frac{d\sigma}{d y_+ d y_- d^2 p_{W\perp}} = \frac{1}{16 \pi^2 {\hat s}^2}
\, x_1 g(x_1,\mu^2) \, x_2 g(x_2,\mu^2)
\overline{ | {\cal M}_{g g \to W^+ W^-}(\hat s, \hat t, \hat u)
  |^2} \, .
\label{gg_fusion}
\end{equation}
This contribution is formally higher order in pQCD than 
the $q \bar q$ annihilation, but may be large numerically at higher energies
when $x_1$ and $x_2$ become very small.

\subsection{$\gamma \gamma \to W^+ W^-$ mechanism}

In this section, we briefly discuss inclusive $\gamma \gamma \to W^+ W^-$
mechanisms. 
We shall calculate that contribution to the inclusive
$p p \to W^+ W^- X$ process for the first time in the literature.

\begin{figure*}
\begin{center}
\includegraphics[width=5cm]{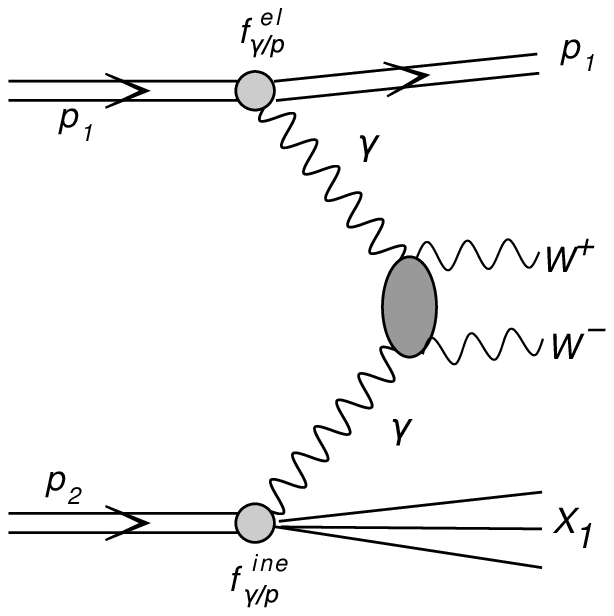}
\includegraphics[width=5cm]{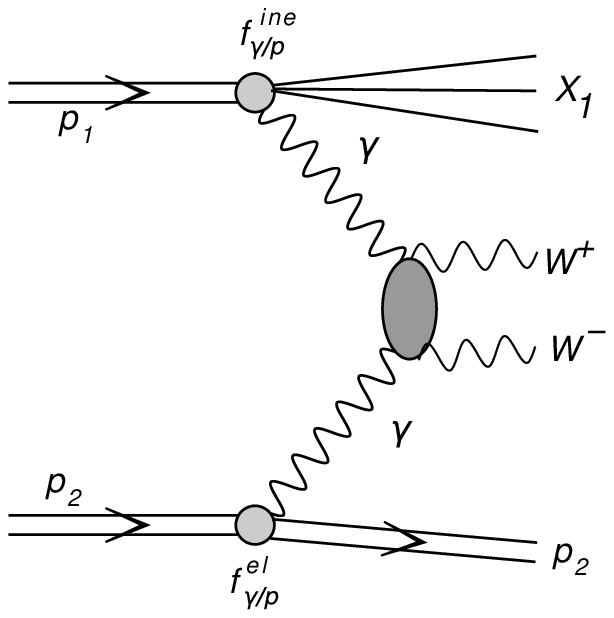}
\includegraphics[width=5cm]{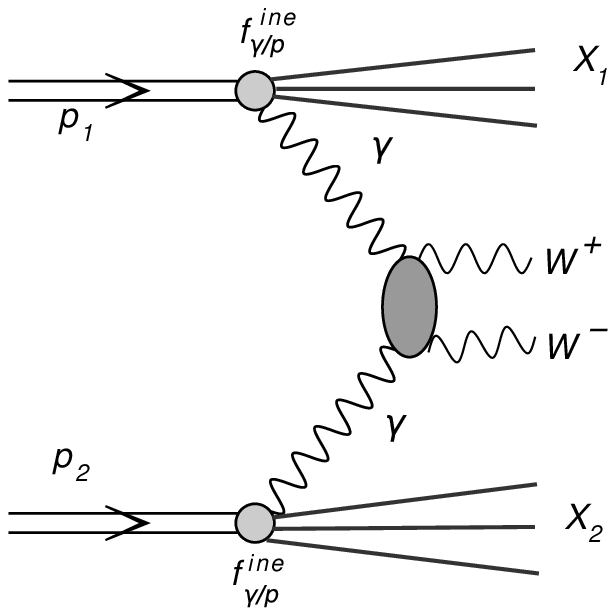}
\caption{Diagrams representing inelastic photon-photon induced mechanisms 
for production of $W^+ W^-$ pairs.
}
\label{fig:new_diagrams}
\end{center}
\end{figure*}

If at least one photon is a constituent of the nucleon then 
the mechanisms presented in Fig.\ref{fig:new_diagrams} are possible.
In these cases at least one of participating proton does not survive 
the $W^+ W^-$ production process. In the following we consider 
two different approaches to the problem.

\subsubsection{Naive approach to photon flux}

Some $\gamma \gamma$ induced processes 
($\gamma \gamma \to H^+ H^-, L^+ L^-$) were discussed long time ago
in Ref.\cite{DGNR94}. In their approach the photon distribution
in the proton is a convolution of the distribution of quarks
in the proton and the distribution of photons in quarks/antiquarks
\begin{equation}
f_{\gamma/p} = f_q \otimes f_{\gamma/q}   \; ,
\end{equation}
which can be written mathematically as
\begin{equation}
x f_{\gamma/p}(x) = \sum_{q} \int_x^1 d x_q f_q(x_q,\mu^2) 
e_q^2 \left( \frac{x}{x_q} \right) f_{\gamma/q}
\left( \frac{x}{x_q},Q_1^2,Q_2^2 \right) \; ,
\label{convolution}
\end{equation}
where the sum runs over all quark and antiquark flavours.
The flux of photons in a quark/antiquark in their approach 
was calculated as:
\begin{equation}
f_{\gamma}(z) = \frac{\alpha_{em}}{2 \pi}
\frac{1 + (1-z)^2}{2} \log \left( \frac{Q_1^2}{Q_2^2} \right) \; .
\label{photon_in_quark}
\end{equation}
The choice of scales in the formulae is a bit ambigous.
They have proposed the following set of scales:
\begin{eqnarray}
Q_1^2 &=& \max ({\hat s}/4-m_W^2, 1 \mbox{GeV}^2) \nonumber \\
Q_2^2 &=& 1 \mbox{GeV}^2 \nonumber \\
\mu^2 &=& {\hat s}/4 \; .
\label{scales}
\end{eqnarray}
We shall try to use the approach here as a reference 
for more refined calculation described in the next subsection.

\subsubsection{MRST-QED parton distributions}

An improved approach how to include photons into inelastic processes 
was proposed some time ago by Martin, Roberts, 
Stirling and Thorne in Ref.\cite{MRST04}. In their approach photon 
is treated on the same footing as quarks, antiquarks and gluons.
Below we repeat the essential points of their formalism which includes 
combined QCD+QED evolution.

They proposed a QED-corrected evolution equations for the parton 
distributions of the proton \cite{MRST04}:
\begin{eqnarray}
{\partial q_i(x,\mu^2) \over \partial \log \mu^2} &=& {\alpha_S\over 2\pi}
\int_x^1 \frac{dy}{y} \Big\{
    P_{q q}(y)\; q_i(\frac{x}{y},\mu^2)
     +  P_{q g}(y)\; g(\frac{x}{y},\mu^2)\Big\}
\, \nonumber \\
&  + &
   {\alpha\over 2\pi} \int_x^1 \frac{dy}{y} \Big\{
    \tilde{P}_{q q}(y)\; e_i^2 q_i(\frac{x}{y},\mu^2)  +  P_{q \gamma}(y)\;
e_i^2 \gamma(\frac{x}{y},\mu^2)         \Big\},  \nonumber \\
{\partial g(x,\mu^2) \over \partial \log \mu^2} &=& {\alpha_S\over 2
\pi} \int_x^1 \frac{dy}{y} \Big\{
    P_{g q}(y)\; \sum_j q_j(\frac{x}{y},\mu^2) 
 + 
    P_{g g}(y)\; g(\frac{x}{y},\mu^2)\Big\},
\, \nonumber \\
   {\partial \gamma(x,\mu^2) \over \partial \log \mu^2}
& =   & {\alpha
\over 2\pi} \int_x^1 \frac{dy}{y} 
   \Big\{ P_{\gamma q}(y)\; \sum_j e_j^2\; q_j(\frac{x}{y},\mu^2) 
+ 
  P_{\gamma \gamma}(y)\; \gamma(\frac{x}{y},\mu^2) \Big\} \; ,
\label{evolution_equations}
\end{eqnarray}
where
\begin{eqnarray}
{\tilde P}_{qq} = C_F^{-1} P_{qq}, & &   P_{\gamma q} = 
C_F^{-1} P_{g q}, \nonumber \\
P_{q\gamma} = T_R^{-1} P_{q g} , & &  P_{\gamma \gamma} = - 
\frac{2}{3}\; \sum_i e_i^2\; \delta(1-y) \nonumber
\end{eqnarray}
The parton distributions in Eq.(\ref{evolution_equations}) fullfil 
the standard momentum sum rule:
\begin{equation}
  \int_0^1 dx\;  x\; \Big\{\sum_i q_i(x,\mu^2) + g(x,\mu^2) + \gamma(x,\mu^2)
     \Big\}  = 1 \; .
\end{equation}
%

\subsubsection{Cross section for photon-photon processes}

In leading order corresponding triple differential cross section 
for inelastic-inelastic photon-photon contribution can be written as
usually in the parton-model formalism:
\begin{eqnarray}
\frac{d \sigma^{\gamma_{in} \gamma_{in}}}{d y_1 d y_2 d^2p_t} &=& \frac{1}{16 \pi^2 {\hat s}^2}
x_1 \gamma_{in}(x_1,\mu^2) \; x_2 \gamma_{in}(x_2,\mu^2) \;
\overline{|{\cal M}_{\gamma \gamma \to W^+W^-}|^2} \; .
\end{eqnarray}

The above contribution includes only cases when both nucleons do not survive
the collision and nucleon debris is produced instead. The case when
nucleon survives the collision has to be considered separately. In this
case one can include corresponding photon distributions where extra "el"
index will be added to denote that physical situation. Corresponding
contributions to the cross section can be then written as:
\begin{eqnarray}
\frac{d \sigma^{\gamma_{in} \gamma_{el}}}{d y_1 d y_2 d^2p_t} &=& \frac{1}{16 \pi^2 {\hat s}^2}
x_1 \gamma_{in}(x_1,\mu^2) \; x_2 \gamma_{el}(x_2,\mu^2) \;
\overline{|{\cal M}_{\gamma \gamma \to W^+W^-}|^2} \; ,\nonumber \\
\frac{d \sigma^{\gamma_{el} \gamma_{in}}}{d y_1 d y_2 d^2p_t} &=& \frac{1}{16 \pi^2 {\hat s}^2}
x_1 \gamma_{el}(x_1,\mu^2) \; x_2 \gamma_{in}(x_2,\mu^2) \;
\overline{|{\cal M}_{\gamma \gamma \to W^+W^-}|^2} \; ,\nonumber \\
\frac{d \sigma^{\gamma_{el} \gamma_{el}}}{d y_1 d y_2 d^2p_t} &=& \frac{1}{16 \pi^2 {\hat s}^2}
x_1 \gamma_{el}(x_1,\mu^2) \; x_2 \gamma_{el}(x_2,\mu^2) \; 
\overline{|{\cal M}_{\gamma \gamma \to W^+W^-}|^2} \; , \\ 
\label{subleading_contributions}
\end{eqnarray}
for inelastic-elastic, elastic-inelastic and elastic-elastic
components, respectively.
In the following the elastic photon fluxes are calculated using the Drees-Zeppenfeld 
parametrization \cite{DZ}, where a simple parametrization of 
nucleon electromagnetic form factors is used.

\subsubsection{Resolved photons}

\begin{figure*}
\begin{center}
\includegraphics[width=4cm]{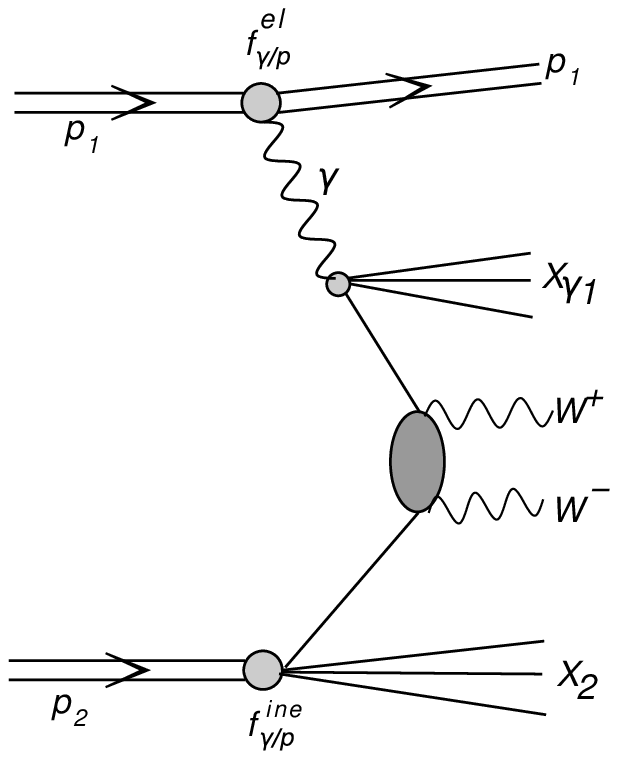}
\includegraphics[width=4cm]{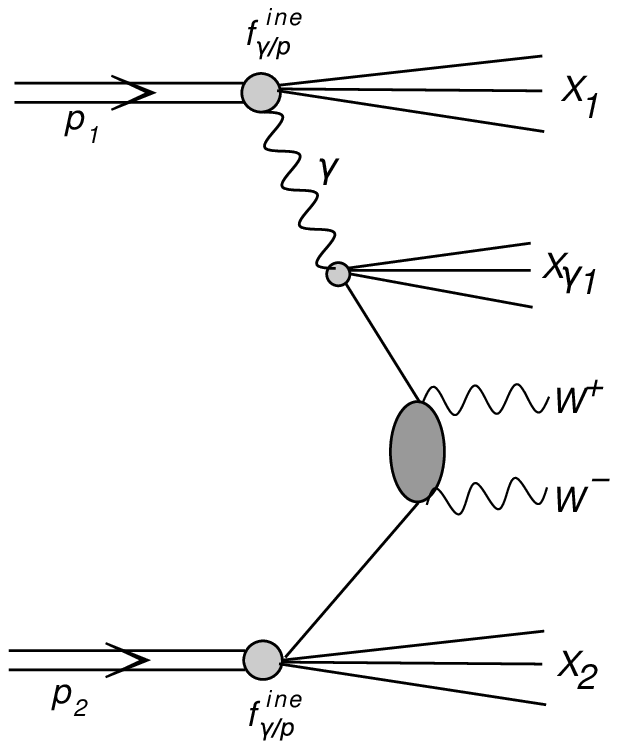}
\includegraphics[width=4cm]{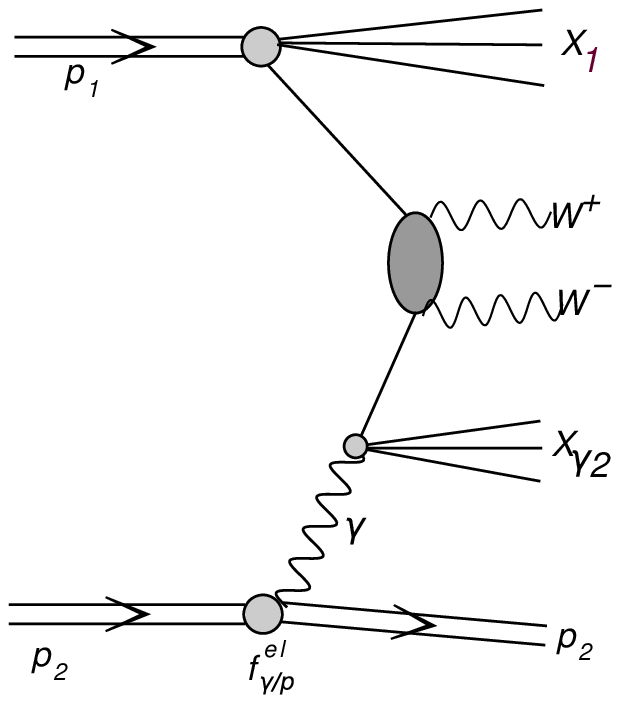}
\includegraphics[width=4cm]{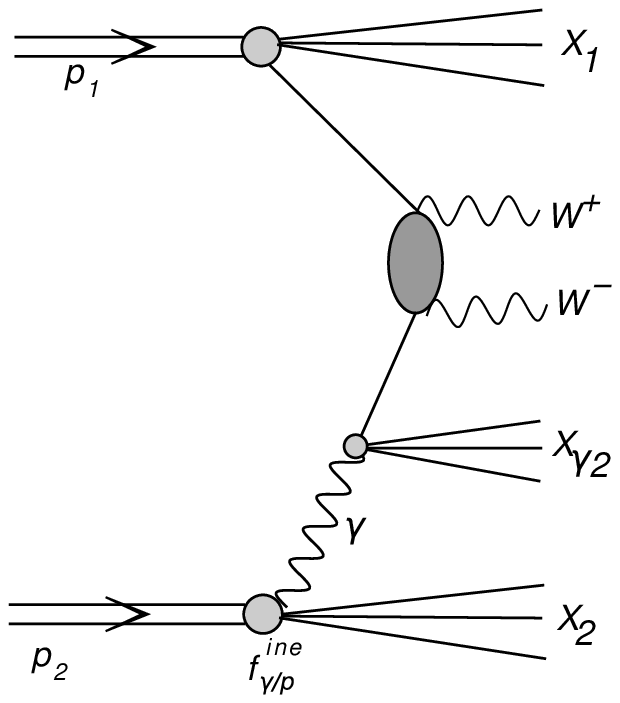}
\caption{Diagrams representing some single resolved photon 
mechanisms with quark-antiquark annihilation for production of $W^+ W^-$
pairs. There exist also similar diagrams with gluon-gluon subprocesses.
}
\label{fig:resolved1}
\end{center}
\end{figure*}

So far we have discussed direct photonic contributions. In general, photon may
also develop its hadronic structure. Then asymmetric diagrams with one photon
attached to the upper or lower proton,
as shown in Fig.\ref{fig:resolved1}, become possible.
Now extra photon remnant debris 
(called $X_{\gamma, 1}$ or $X_{\gamma, 2}$ in the figure) 
appears in addition. One may expect that such diagrams lead to quite
asymmetric distributions in rapidity of $W$ bosons with maxima in 
forward and/or backward directions.

\begin{figure*}
\begin{center}
\includegraphics[width=4cm]{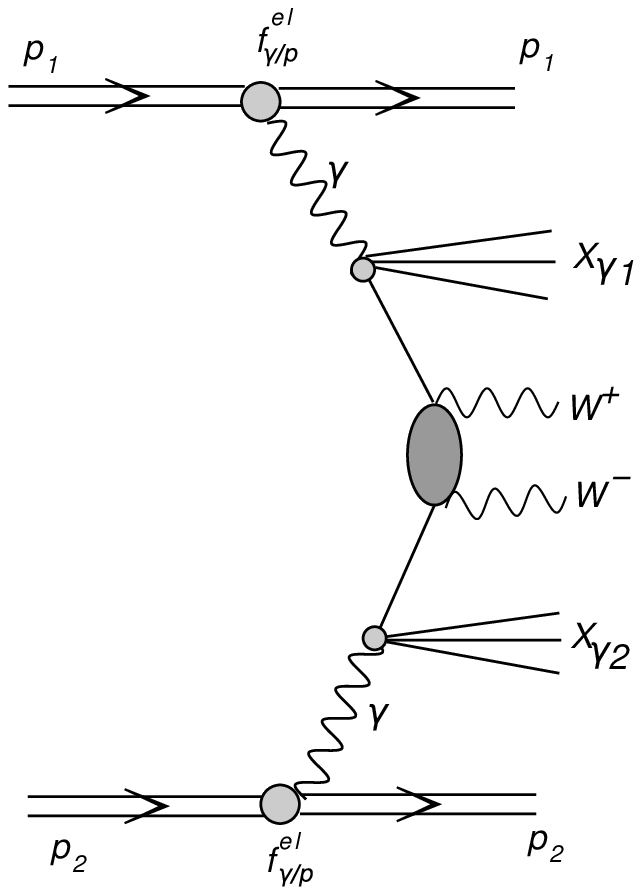}
\includegraphics[width=4cm]{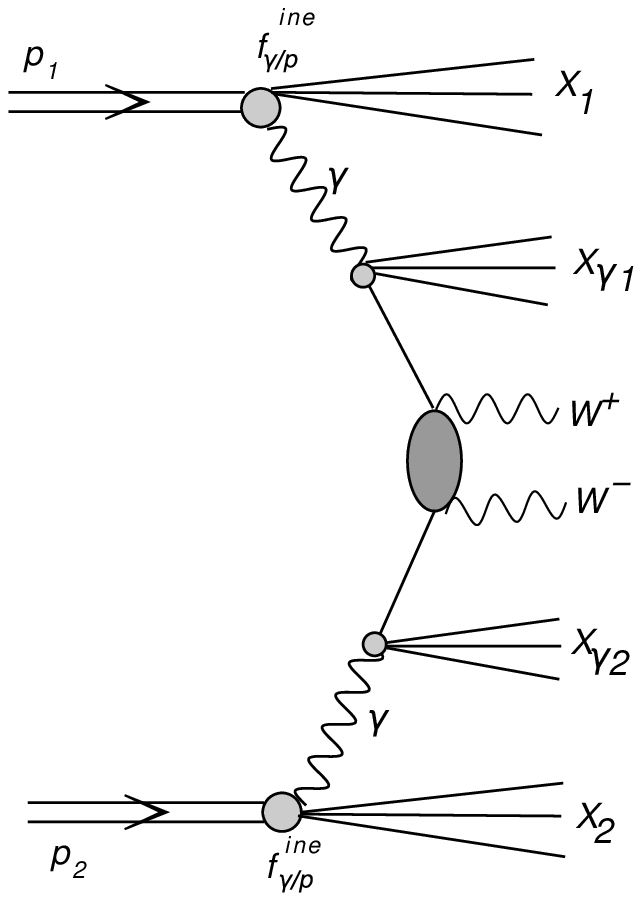}
\includegraphics[width=4cm]{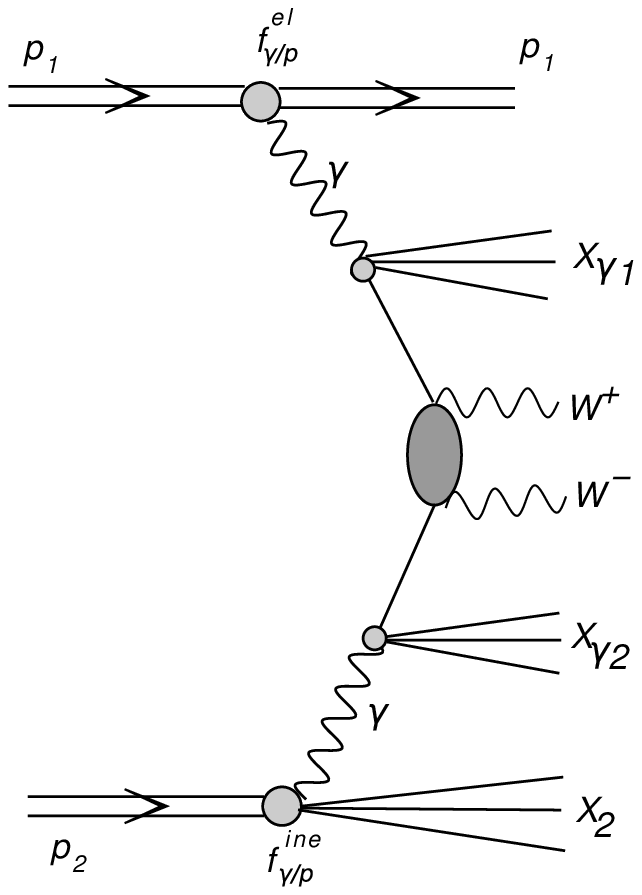}
\includegraphics[width=4cm]{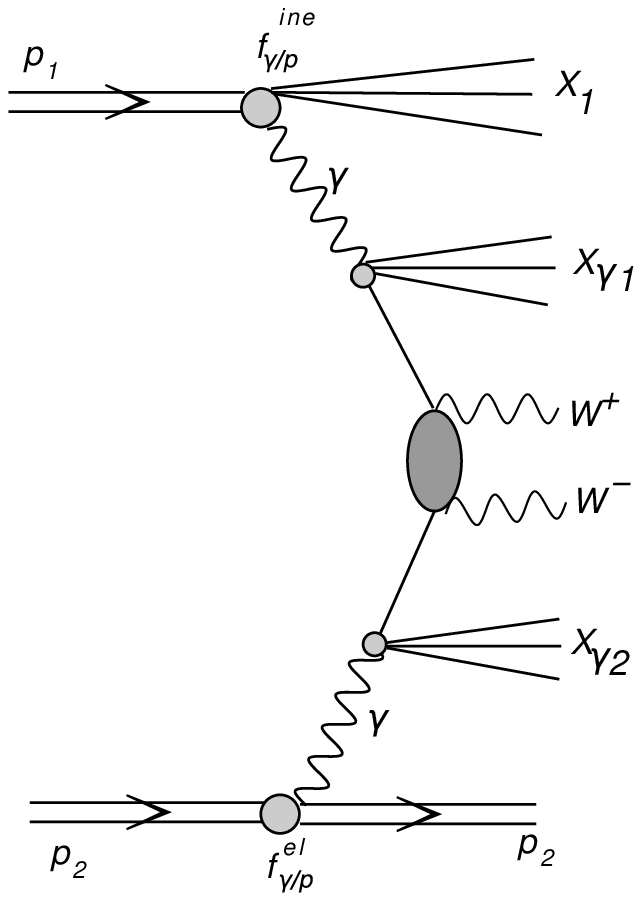}
\caption{Diagrams representing mechanisms with two resolved photons 
for production of $W^+ W^-$ pairs.
}
\label{fig:resolved2}
\end{center}
\end{figure*}

Another type of diagrams with resolved photons is shown in
Fig.\ref{fig:resolved2}. We expect the contributions of the second
category of diagrams, shown in Fig.\ref{fig:resolved2}, to be rather
small. 

In the case of resolved photons, the ``photonic'' quark/antiquark 
distributions in a proton must be calculated first. This can be done 
as the convolution
\begin{equation}
f_{q/p}^{\gamma} = f_{\gamma/p} \otimes f_{q/\gamma}
\end{equation}
which mathematically means:
\begin{equation}
x f_{q/p}^{\gamma}(x) = \int_x^1 d x_{\gamma} f_{\gamma/p}(x_{\gamma},\mu_s^2) 
\left( \frac{x}{x_{\gamma}} \right) 
f \left( \frac{x}{x_{\gamma}}, \mu_h^2 \right)  \; . 
\label{convolution_resolved}       
\end{equation}
Technically first $f_{\gamma/p}$ in the proton is prepared on a dense
grid for $\mu_s^2 \sim$ 1 GeV$^2$ (virtuality of the photon) and then is used 
in the convolution formula (\ref{convolution_resolved}). 
The second scale is evidently hard $\mu_h^2 \sim M_{WW}^2$.
The result strongly depends on the choice
of the soft scale $\mu_s^2$. In this sense our calculations
are not very precise and must be treated rather as a rough estimate.
The new quark/antiquark distributions of photonic origin are 
used to calculate cross section as for the standard quark-antiquark 
annihilation subprocess.

\subsection{Single diffractive production of $W^+ W^-$ pairs}

\begin{figure*}
\begin{center}
\includegraphics[width=5cm]{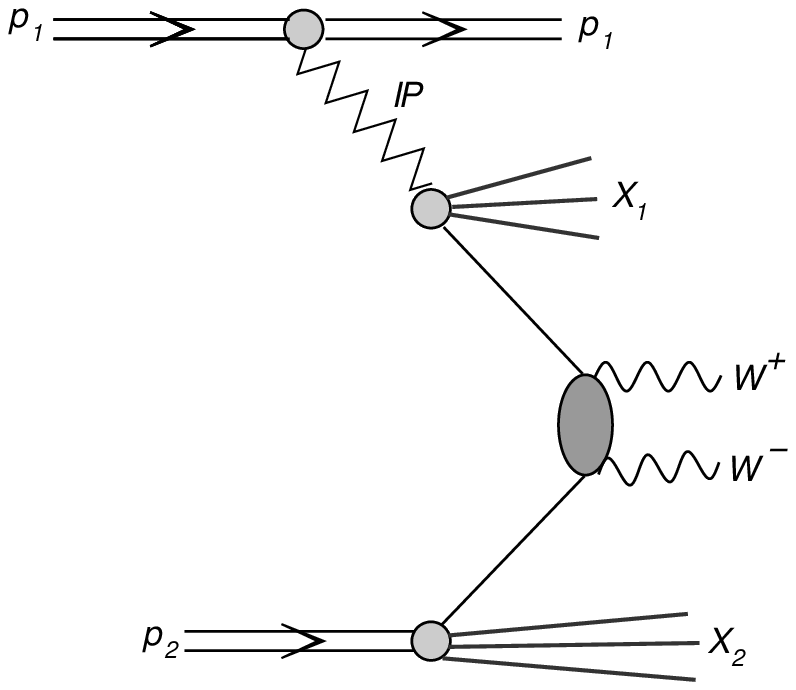}
\includegraphics[width=5cm]{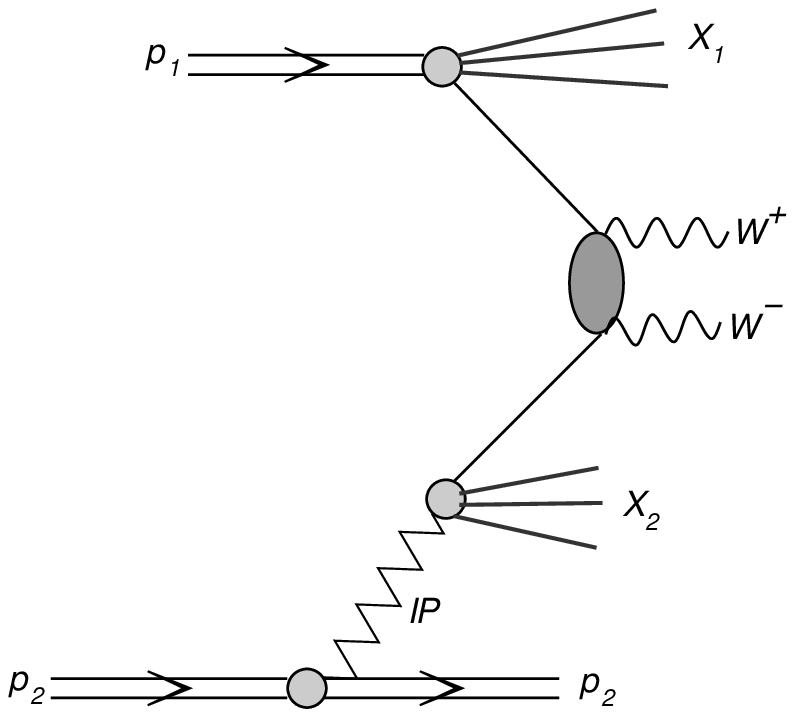}
\caption{Diagrams representing single diffractive mechanism
of production of $W^+ W^-$ pairs.
}
\label{fig:single_diffractive}
\end{center}
\end{figure*}

In the following we apply the resolved pomeron approach 
\cite{IS_ee,IS_ccbar}.
In this approach one assumes that the Pomeron has a
well defined partonic structure, and that the hard process
takes place in a Pomeron--proton or proton--Pomeron (single diffraction) 
or Pomeron--Pomeron (central diffraction) processes.
The mechanism of single diffractive production of $W^+ W^-$ pairs
is shown in Fig.\ref{fig:single_diffractive}.
We calculate triple differential distributions
\begin{eqnarray}
{d \sigma_{SD}^{(1)} \over dy_{1} dy_{2} dp_{t}^2} = {\Big| M \Big|^2 \over 16 \pi^2 \hat{s}^2} 
\,\Big [\, \Big( x_1 q_f^D(x_1,\mu^2) 
\, x_2 \bar q_f(x_2,\mu^2) \Big) \, 
+ \Big( x_1 \bar q_f^D(x_1,\mu^2)
\, x_2  q_f(x_2,\mu^2) \Big) \, \Big ] ,
\nonumber \\ 
\label{DY}
\end{eqnarray}
\begin{eqnarray}
{d \sigma_{SD}^{(2)} \over dy_{1} dy_{2} dp_{t}^2} = {\Big| M \Big|^2 \over 16 \pi^2 \hat{s}^2} 
\,\Big [\, \Big( x_1 q_f(x_1,\mu^2) 
\, x_2 \bar q_f^D(x_2,\mu^2) \Big) \, 
+ \Big( x_1 \bar q_f(x_1,\mu^2)
\, x_2  q_f^D(x_2,\mu^2) \Big) \, \Big ] ,
\nonumber \\ 
\label{SD}
\end{eqnarray}
\begin{eqnarray}
{d \sigma_{CD} \over dy_{1} dy_{2} dp_{t}^2} =  {\Big| M \Big|^2 \over 16 \pi^2 \hat{s}^2} 
\,\Big [\, \Big( x_1 q_f^D(x_1,\mu^2) 
\, x_2 \bar q_f^D(x_2,\mu^2) \Big) \, 
+ \Big( x_1 \bar q_f^D(x_1,\mu^2)
\, x_2  q_f^D(x_2,\mu^2) \Big) \,\Big ] 
\nonumber \\ 
\label{DD}
\end{eqnarray}
for single-diffractive and central-diffractive production, 
respectively.
The matrix element squared for the $q \bar q \to W^{+} W^{-}$
process is the same as previously discussed for the nondiffractive processes.

In this approach longitudinal momentum fractions are calculated as
\begin{eqnarray}
x_1 = {m_{t} \over \sqrt{s}}  \Big( e^{y_{2}} + e^{y_{2}} \Big) , \\
x_2 = {m_{t} \over \sqrt{s}}  \Big( e^{-y_{1}} + e^{-y_{2}} \Big)
\nonumber
\end{eqnarray}
with $m_{t} = \sqrt{ (p_{t}^2 + m_{W}^2)} \approx p_{t} $.
The distribution in the $WW$ invariant mass can be 
obtained by binning differential cross section in $M_{WW}$.

In the present analysis we consequently do not calculate higher-order 
contributions. In principle, they could be included effectively with 
the help of a so-called $K$-factor.

The 'diffractive' quark/antiquark distribution of
flavour $f$ can be obtained by a convolution of the flux of Pomerons
$f_\Pom(x_\Pom)$ and the parton distribution in the Pomeron 
$q_{f/\Pom}(\beta, \mu^2)$:
\begin{eqnarray}
q_f^D(x,\mu^2) = \int d x_\Pom d\beta \, \delta(x-x_\Pom \beta) 
q_{f/\Pom} (\beta,\mu^2) \, f_\Pom(x_\Pom) \, 
= \int_x^1 {d x_\Pom \over x_\Pom} \, f_\Pom(x_\Pom)  
q_{f/\Pom}({x \over x_\Pom}, \mu^2) \, . \nonumber \\
\label{diffractive_convolution}
\end{eqnarray}
The flux of Pomerons $f_\Pom(x_\Pom)$ enters in the form integrated over 
four--momentum transfer 
\begin{eqnarray}
f_\Pom(x_\Pom) = \int_{t_{min}}^{t_{max}} dt \, f(x_\Pom,t) \, ,
\label{flux_of_Pom}
\end{eqnarray}
with $t_{min}, t_{max}$ being kinematic boundaries.

Both pomeron flux factors $f_{\Pom}(x_{\Pom},t)$ as well 
as quark/antiquark distributions in the pomeron are taken from 
the H1 collaboration analysis of diffractive structure function
and diffractive dijets at HERA \cite{H1}. 
The factorization scale for diffractive parton distributions is taken as
$\mu^2 = M_{WW}^2$.

In the present analysis we consider both pomeron and subleading reggeon
contributions. The corresponding diffractive quark distributions
are obtained by replacing pomeron flux by the reggeon flux and
quark/antiquark distributions in the pomeron by their counterparts
in subleading reggeon(s). The other details can be found in \cite{H1}.
In the case of pomeron exchange the upper limit in 
(\ref{diffractive_convolution}) is 0.1 and for reggeon exchange 0.2.
In our opinion, the whole Regge formalism does not apply above these limits.

Up to now we have assumed Regge factorization which is known
to be violated in hadron-hadron collisions.
It is known that these are soft interactions which lead to an extra 
production of particles which fill in the rapidity gaps related 
to pomeron exchange.

Different models of absorption corrections 
(one-, two- or three-channel approaches) 
for diffractive processes were presented in the literature.
The absorption effects for the diffractive processes were calculated e.g.
in Ref \cite{Khoze,Maor}.
The different models give slightly different predictions.
Usually an average value of the gap survival probability
$<|S|^2>$ is calculated first and then the cross sections for different
processes is multiplied by this value.
We shall follow this somewhat simplified approach also here.
Numerical values of the gap survival probability can be found 
in \cite{Khoze,Maor}.
The survival probability depends on the collision energy.
It is sometimes parametrized as:
\begin{equation}
<|S|^2>(\sqrt{s}) = \frac{a}{b+\ln(\sqrt{s})} \; .
\end{equation}
The numerical values of the parameters can be found in original
publications.
In general, the absorptive corrections for single and 
central diffractive processes are somewhat different.

\subsection{Double parton scattering}

\begin{figure*}
\begin{center}
\includegraphics[width=5cm]{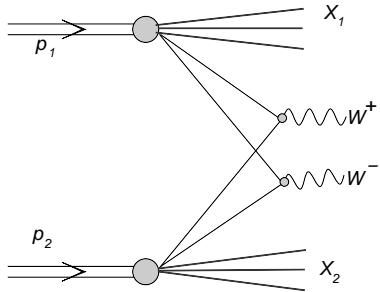}
\caption{Diagram representing double parton scattering mechanism
of production of $W^+ W^-$ pairs.
}
\label{fig:DPS}
\end{center}
\end{figure*}

The diagram representating the double parton scattering
process is shown in Fig.\ref{fig:DPS}.

The cross section for double parton scattering is often modelled
in the factorized anzatz which in our case would mean:
\begin{equation}
\sigma_{W^+ W^-}^{DPS} = \frac{1}{\sigma_{qq}^{eff}} 
\sigma_{W^{+}}
\sigma_{W^{-}}
\; .
\label{factorized_model}
\end{equation}
In general, the parameter $\sigma_{q q}$ does not need to be the same
as for gluon-gluon initiated processes $\sigma_{gg}^{eff}$.
In the present, rather conservative, calculations we take it to be
$\sigma_{qq}^{eff} = \sigma_{gg}^{eff}$ = 15 mb.
The latter value is known within about 10 \% from 
systematics of gluon-gluon initiated processes at the Tevatron and LHC.

The factorized model (\ref{factorized_model}) can be generalized 
to more differential distributions (see e.g. \cite{LMS2012,MS2013}).
For example in our case of $W^{+} W^{-}$ production the cross section
differential in $W$ boson rapidities can be written as:
\begin{equation}
\frac{d \sigma_{W^+ W^-}^{DPS}}{d y_{+} d y_{-}} =
\frac{1}{\sigma_{qq}^{eff}} 
\frac{d\sigma_W^{+}}{d y_{+}}
\frac{d\sigma_W^{-}}{d y_{-}} \; .
\label{generalized_factorized_model}
\end{equation}
In particular, in leading-order approximation the cross section for 
quark-antiquark annihilation reads:
\begin{equation}
\frac{d\sigma}{dy} = \sum_{ij} 
\left( 
  x_1 q_{i/1}(x_1,\mu^2) x_2 {\bar q}_{j/2}(x_2,\mu^2) 
+ x_1 {\bar q}_{i/1}(x_1,\mu^2) x_2 q_{j/2}(x_1,\mu^2) \right)
\overline{|{\cal M}_{ij \to W^{\pm}}|^2} \; ,
\label{rapidity_of_W}
\end{equation}
where the matrix element for quark-antiquark annihilation to $W$ bosons
(${\cal M}_{ij \to W^{\pm}}$) contains Cabibbo-Kobayashi-Maskawa 
matrix elements.
In the present paper for illustration we shall show the 
$\frac{d\sigma}{dy_{+} dy_{-}}$, as well as distributions
in rapidity distance between $W^{+}$ and $W^{-}$ and distribution
in $M_{WW}$.
In the approximations made here (leading order approximation, no
transverse momenta of W bosons)
\begin{equation}
M_{WW}^2 = 2 M_W^2 \left( 1 + \mbox{cosh}(y_1-y_2) \right) \; .
\label{invariant_mass}
\end{equation}

When calculating the cross section for single $W$ boson production
in leading-order approximation a well known Drell-Yan $K$-factor can be 
included. The double-parton scattering would be then multiplied by $K^2$.

\section{Results}

Before a detailed survey of results of the different contributions discussed
in the present paper let us concentrate on
some technical details concerning inelastic photon-photon contributions.
In Fig.\ref{fig:dsig_dy_approx} we show rapidity distributions for 
the naive (left panel) and QCD improved (right panel) approaches
discussed in section \ref{sec:inclusive}.
While in the naive approximation the elastic-elastic component
is the largest and inelastic-inelastic is the smallest, in 
the QCD improved approach the situation is reversed. Here in the QCD 
improved calculations $\mu_F^2 = M_{WW}^2$ was used as 
the factorization scale. 

\begin{figure}
\begin{center}
\includegraphics[width=8cm]{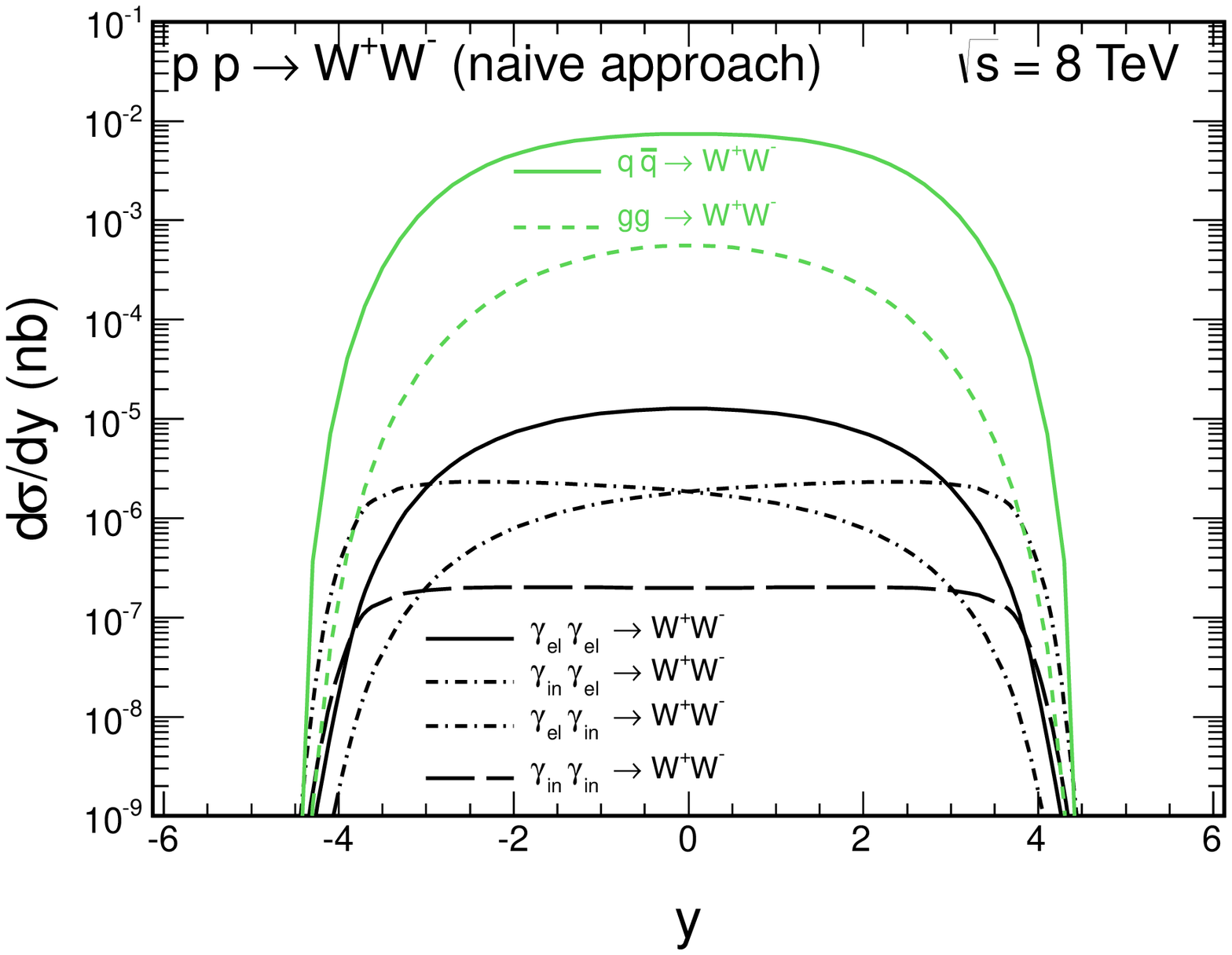}
\includegraphics[width=8cm]{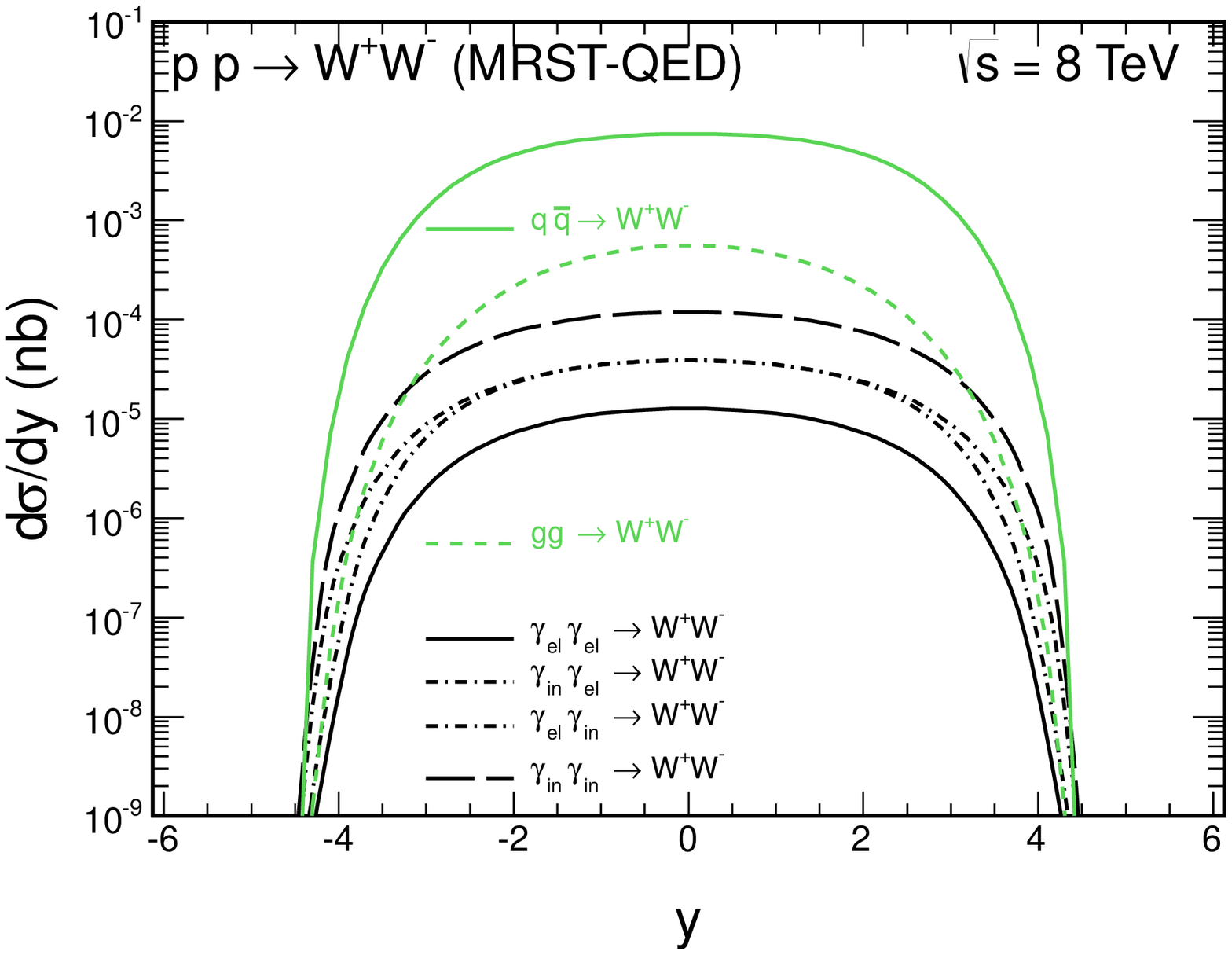}
\end{center}
\caption {Rapidity distribution of $W$ bosons for $\sqrt{s}$ = 8 TeV.
The left panel shows results for the naive approach often used
in the past, while the right panel shows the result with
the QCD improved method proposed in Ref.\cite{MRST04}.
}
\label{fig:dsig_dy_approx}
\end{figure}
\begin{figure}
\begin{center}
\includegraphics[width=8cm]{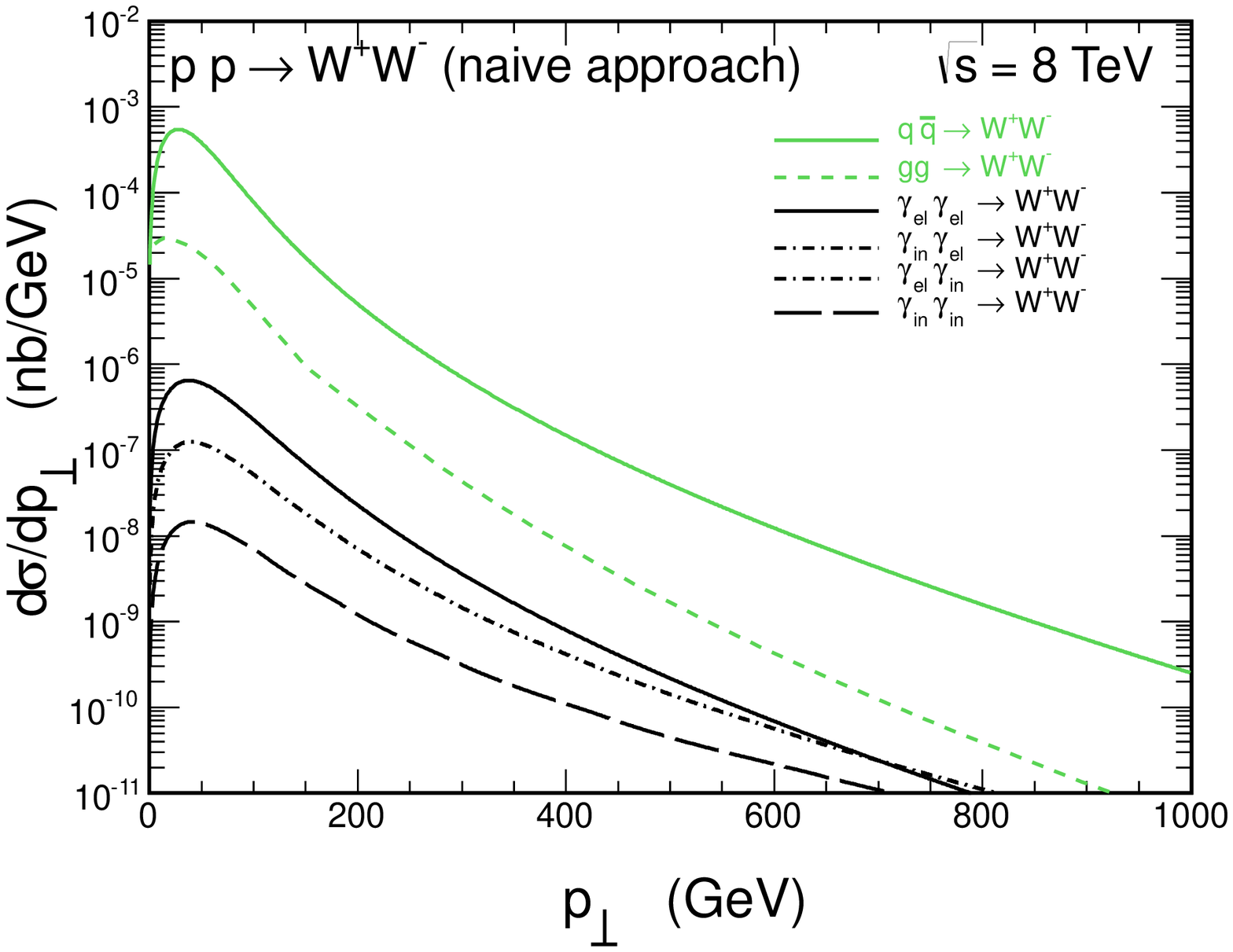}
\includegraphics[width=8cm]{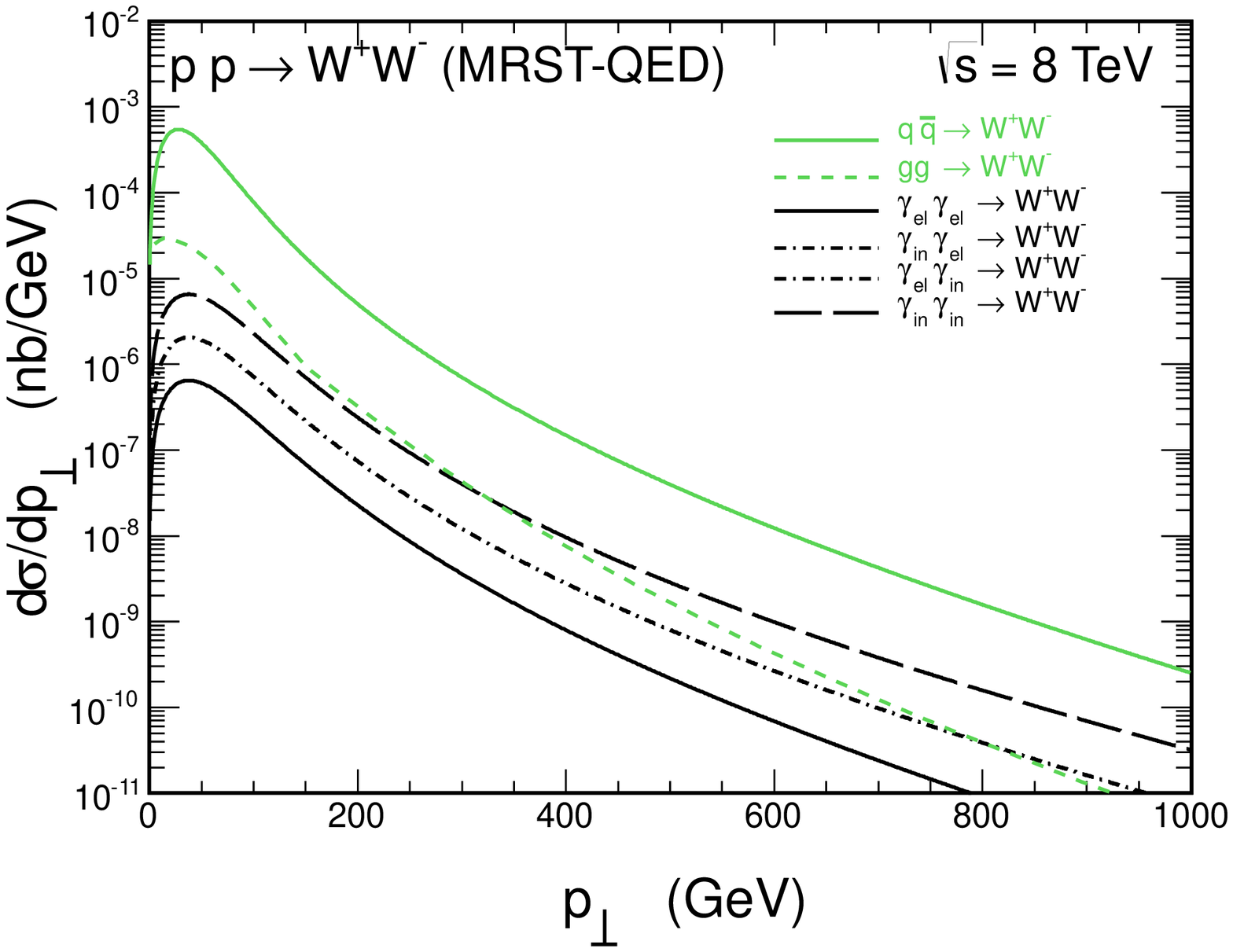}
\end{center}
\caption {Transverse momentum distribution of $W$ bosons for 
$\sqrt{s}$ = 8 TeV.
The left panel shows results for the naive approach often used
in the past, while the right panel shows the result with
the QCD improved method proposed in Ref.\cite{MRST04}.
}
\label{fig:dsig_dy_approx}
\end{figure}

\begin{figure}
\begin{center}
\includegraphics[width=8cm]{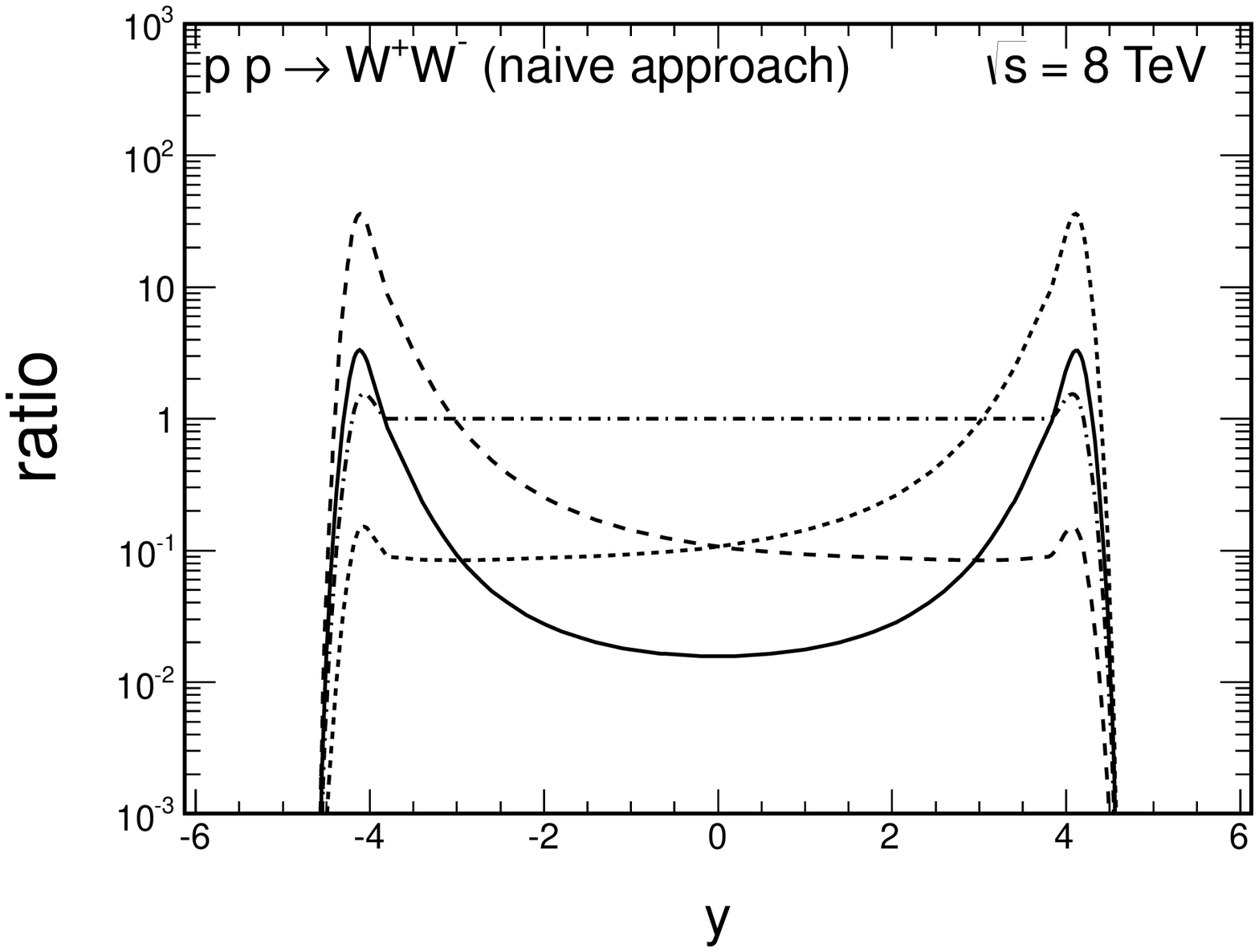}
\includegraphics[width=8cm]{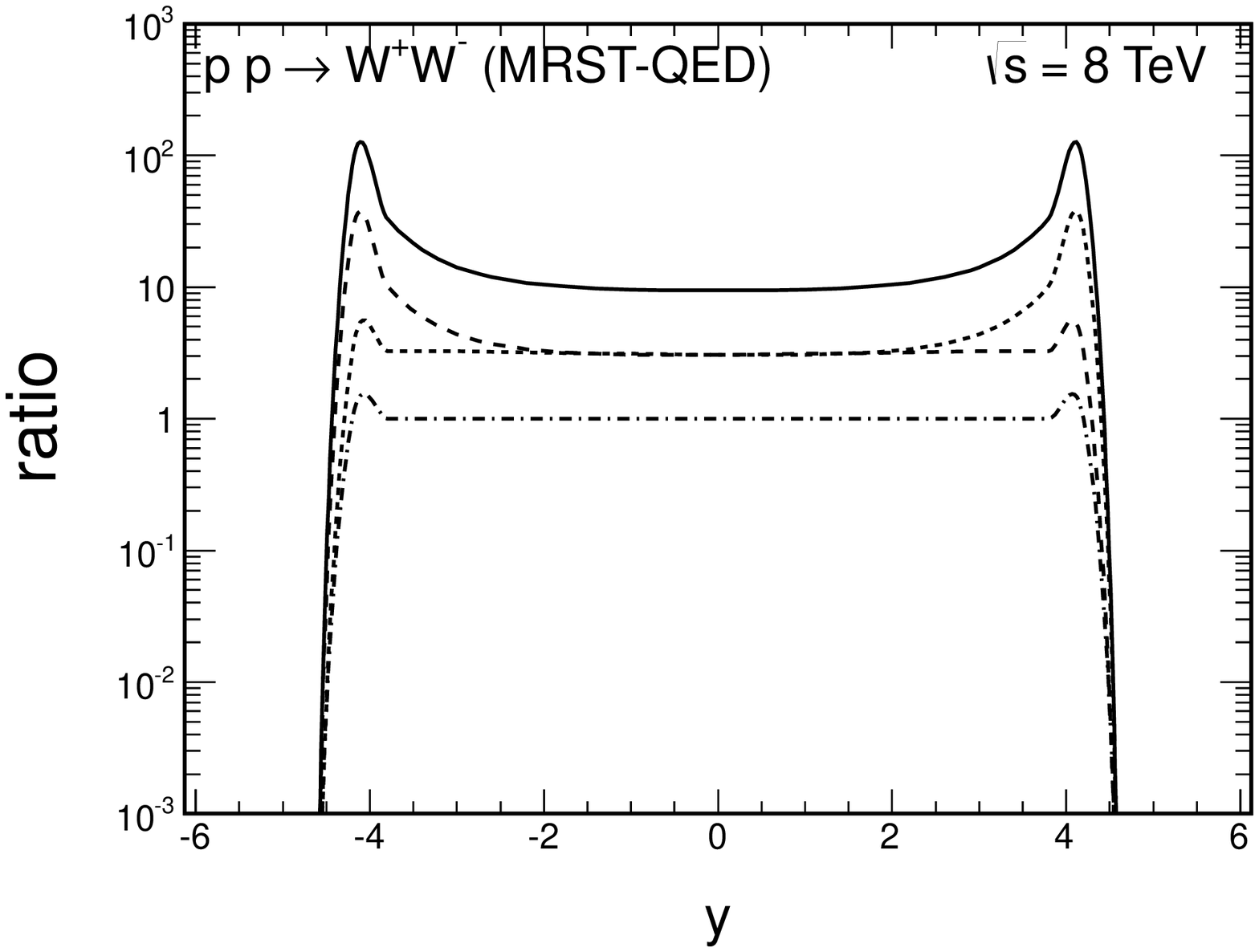}
\end{center}
\caption {The ratio of elastic-inelastic (dashed), inelastic-elastic
  (dotted), inelastic-inelastic (dash-dotted) and all inelastic (solid)
to the elastic-elastic cross section as a function of $W$ boson
rapidity.
}
\label{fig:ratio_y}
\end{figure}

\begin{figure}
\begin{center}
\includegraphics[width=8cm]{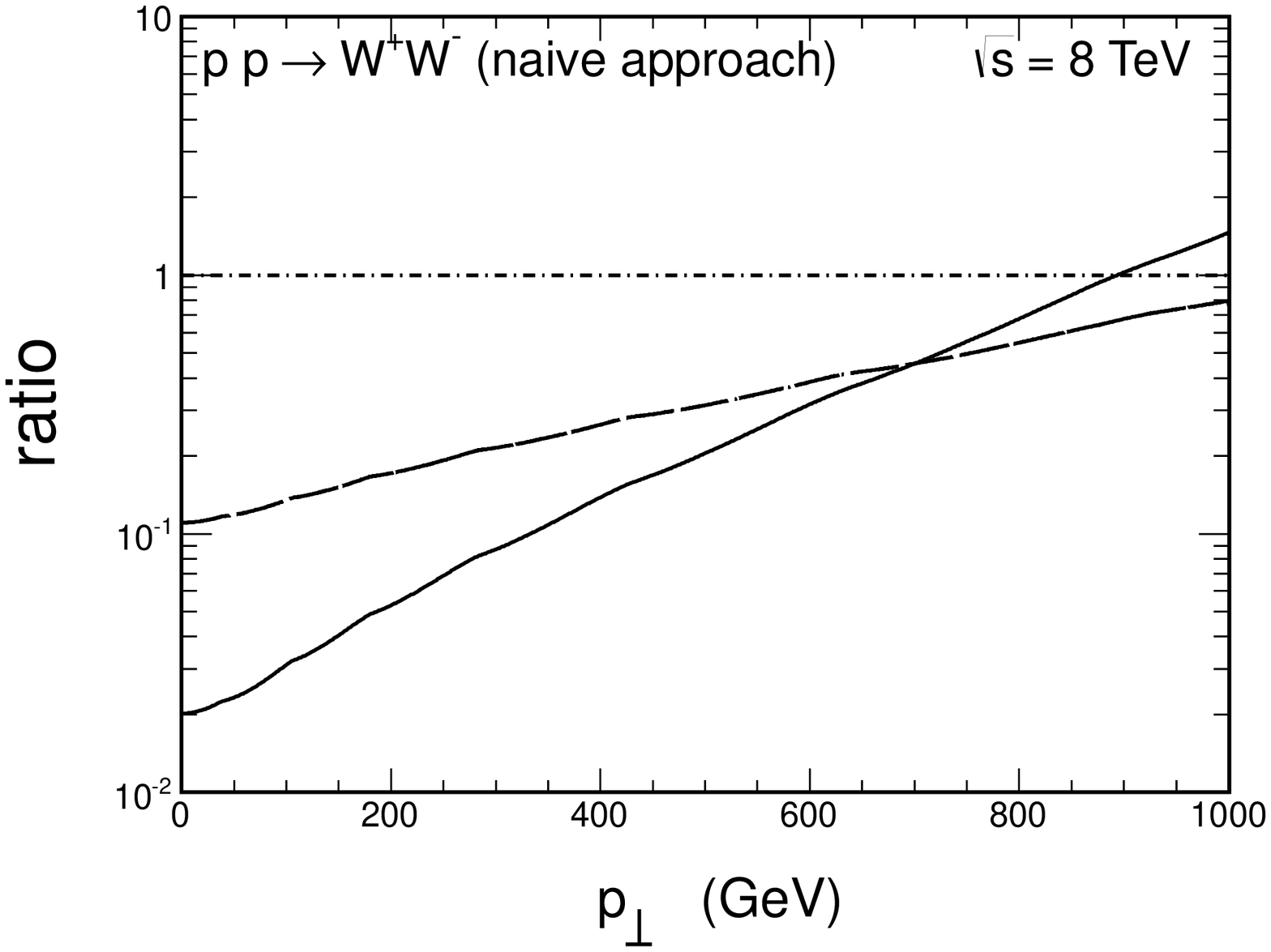}
\includegraphics[width=8cm]{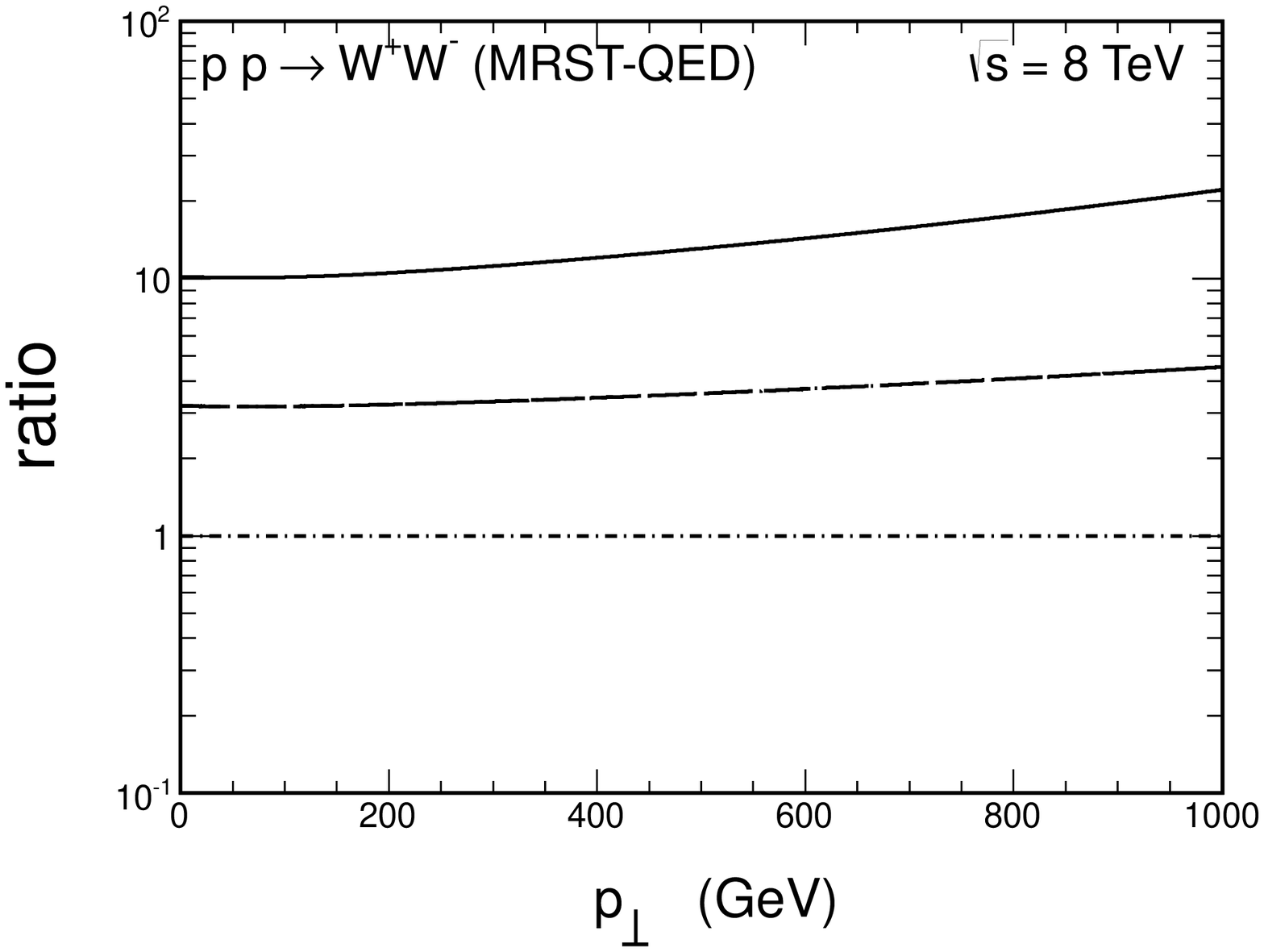}
\end{center}
\caption {The ratio of elastic-inelastic (dashed), inelastic-elastic
  (dotted), inelastic-inelastic (dash-dotted) and all inelastic (solid)
to the elastic-elastic cross section as a function of $W$ boson
transverse momentum.
}
\label{fig:ratio_pt}
\end{figure}

For the searches of anomalous $\gamma \gamma W W$ coupling the ratios
of inelastic-inelastic, elastic-inelastic and inelastic-elastic 
to elastic-elastic one are very useful.
In Fig.\ref{fig:ratio_y} and \ref{fig:ratio_pt} we show such ratios in
rapidity and transverse momentum of $W$ bosons. In the naive approach
the ratios are smaller than 1 except of some small corners of the phase-space.
In the QCD improved approach the ratios become much bigger.
In particular, the ratio for inelastic-inelastic contribution is order of
magnitude larger than 1.

Now we wish to present a systematic survey of all the contributions 
discussed in the present paper. 

\begin{figure}
\begin{center}
\includegraphics[width=8cm]{fig_10b.eps}\\
\includegraphics[width=8cm]{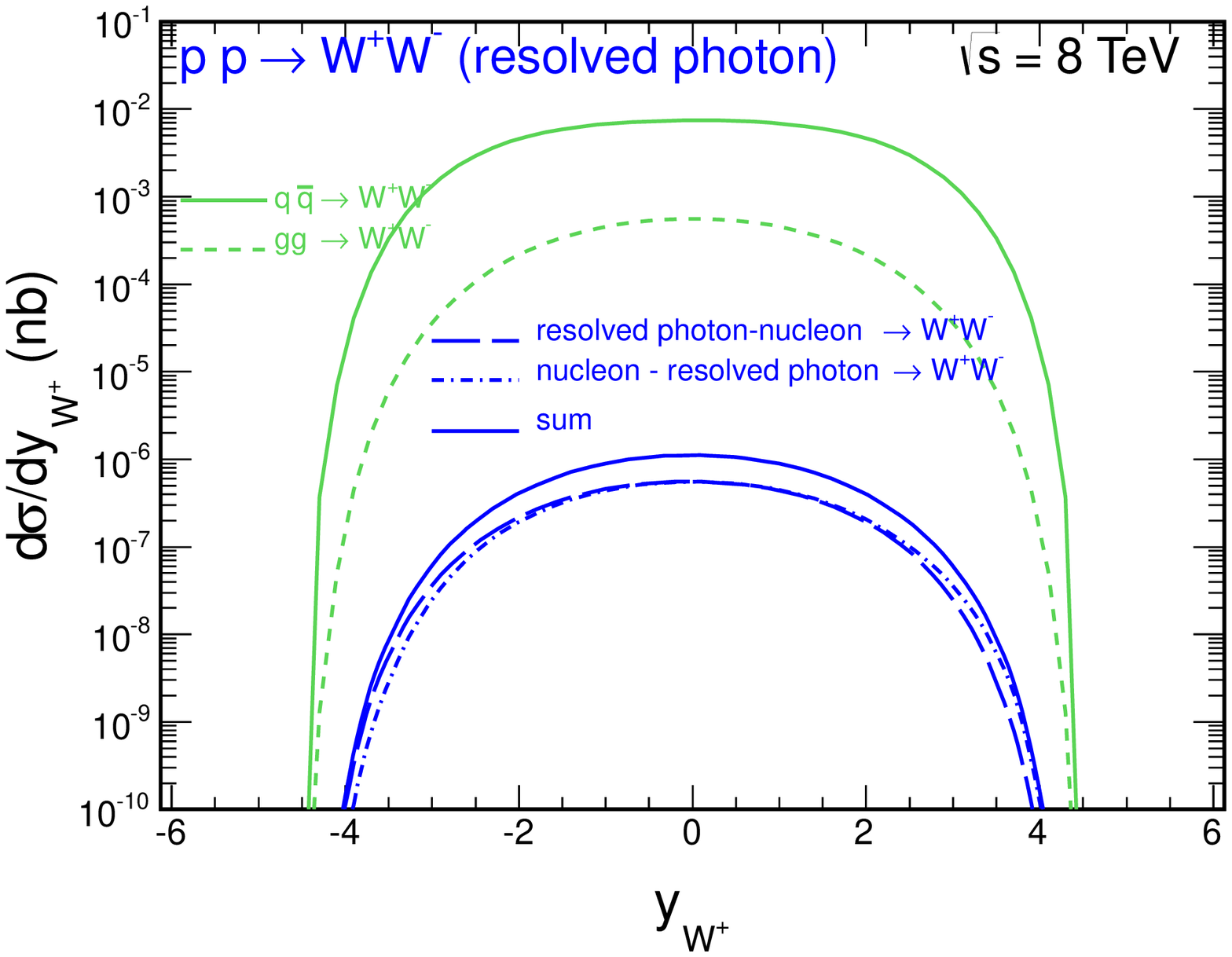}
\includegraphics[width=8cm]{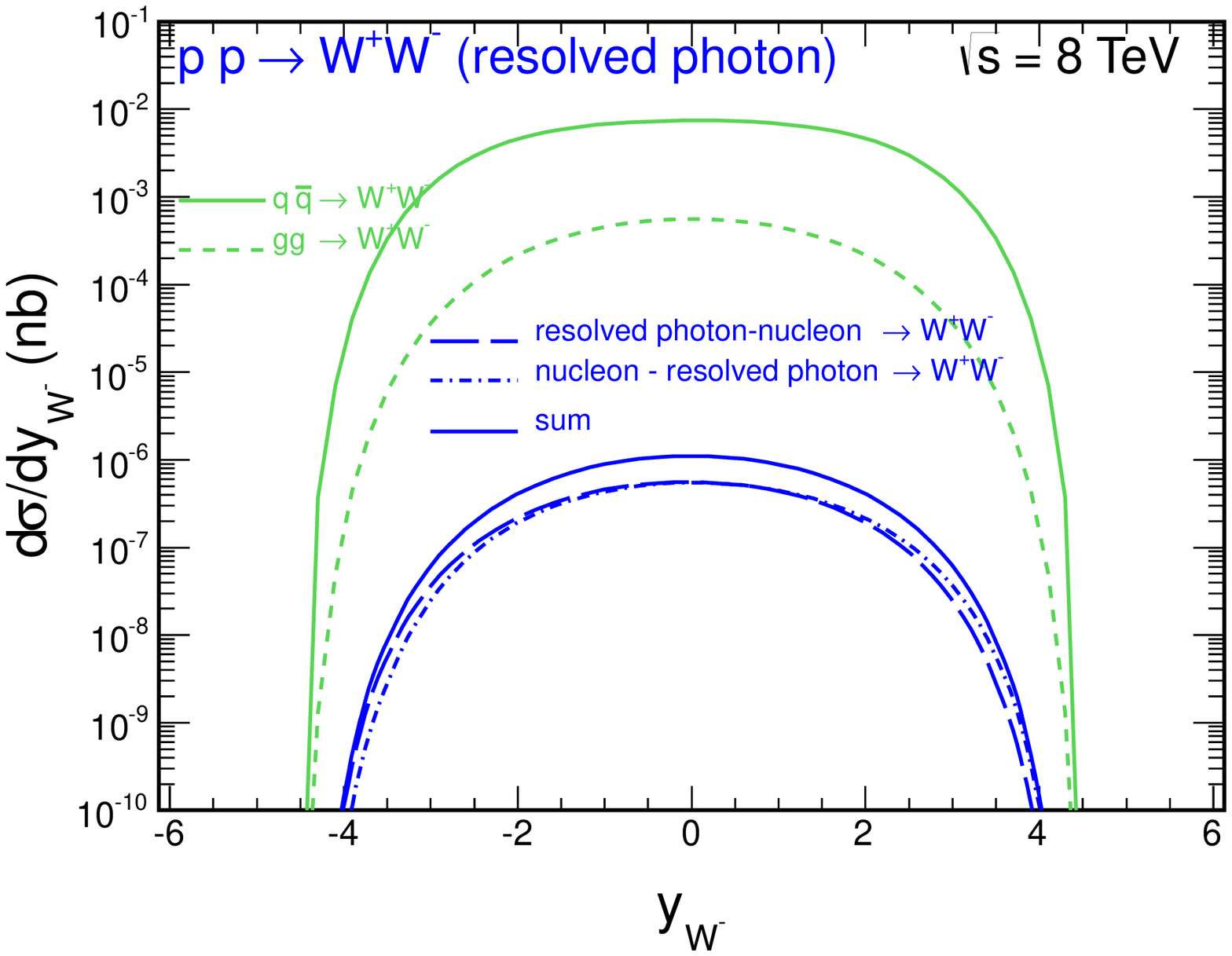}\\
\includegraphics[width=8cm]{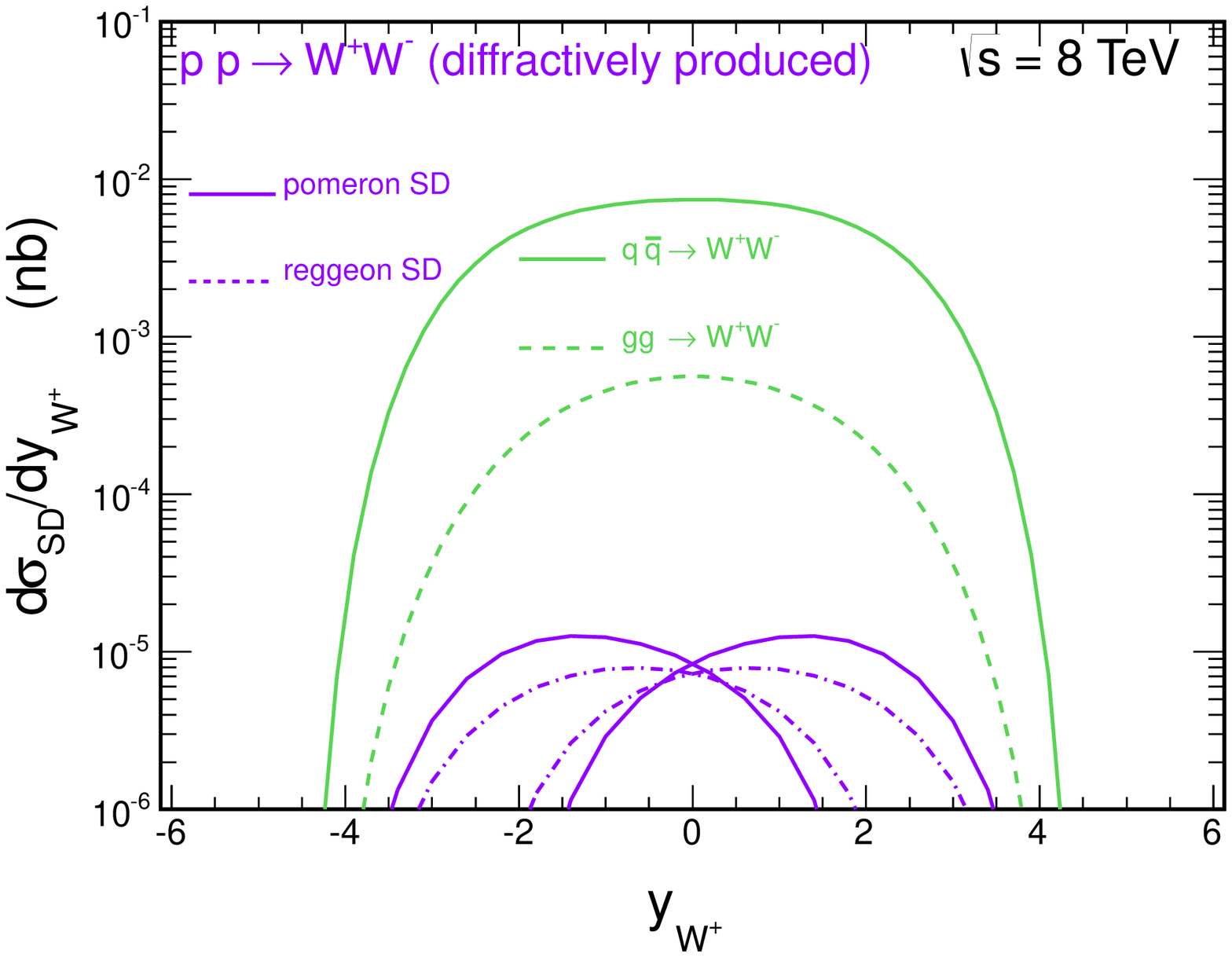}
\includegraphics[width=8cm]{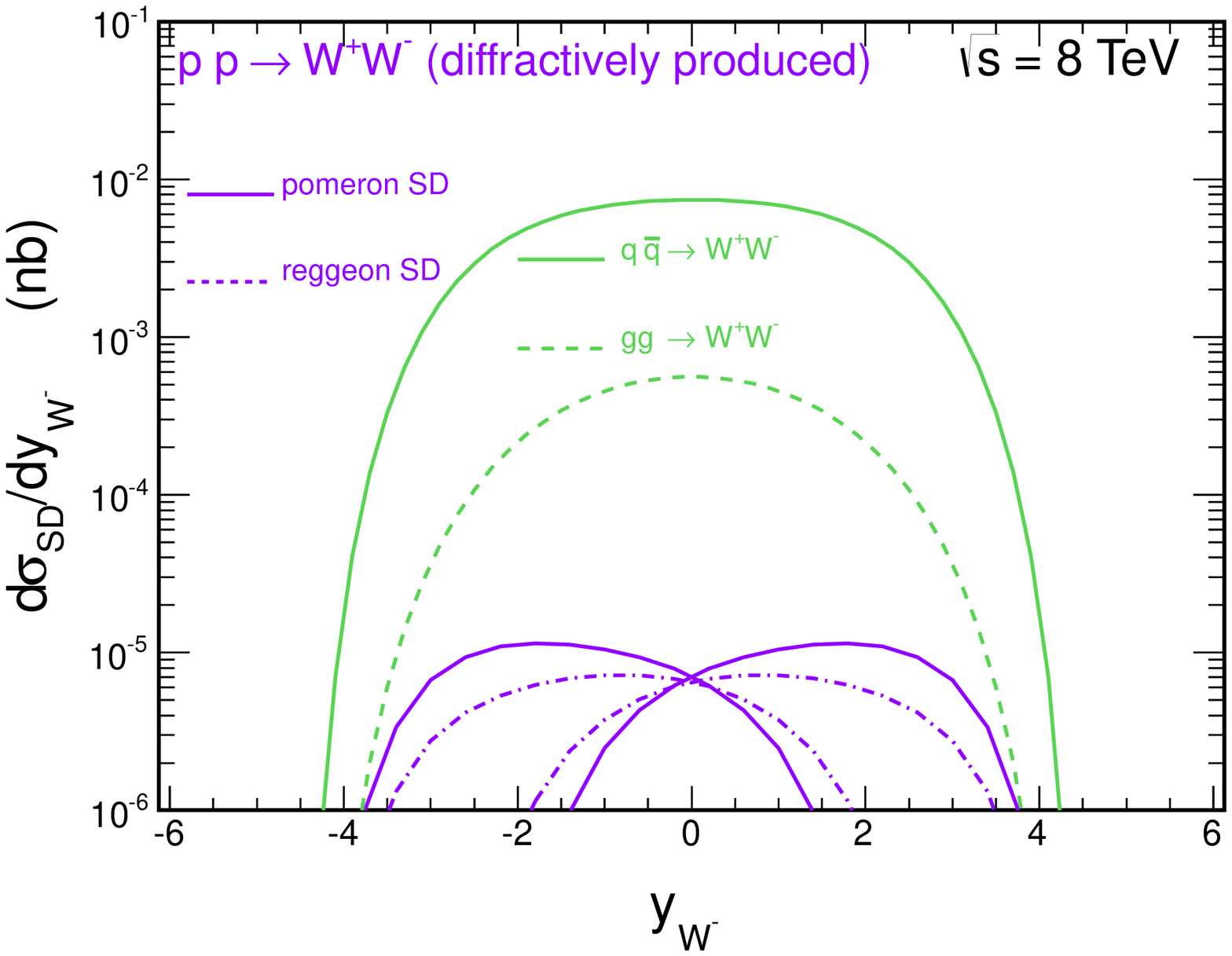}
\end{center}
\caption {Rapidity distribution of $W$ bosons for $\sqrt{s}$ = 8 TeV.
The top panel shows contributions of all photon-photon induced processes, the
middle panels resolved photon contributions and the bottom panels 
distributions of the diffractive contribution. The diffractive
cross section has been multiplied by the gap survival factor
$S_G^2$ = 0.03.
}
\label{fig:dsig_dy}
\end{figure}

The distribution in $W$ boson rapidity is shown in
Fig.\ref{fig:dsig_dy}.
We show separate contributions discussed in the present paper.
The diffractive contribution is an order of magnitude larger than
the resolved photon contribution. The estimated reggeon contribution is
of similar size as the pomeron contribution.
The distributions of $W^+$ and $W^-$ for double-parton scattering 
contribution are different and reflect distribution and, in 
the approximation discussed here, 
have shapes identical as for single production of $W^+$ and $W^-$, 
respectively.
It would be therefore interesting to obtain separate distributions
for $W^+$ and $W^-$ experimentally. This is, however, a rather difficult task.
Distributions of charged leptons (electrons, muons) could also be
interesting in this context.

\begin{figure}
\begin{center}
\includegraphics[width=8cm]{fig_11b.eps}
\includegraphics[width=8cm]{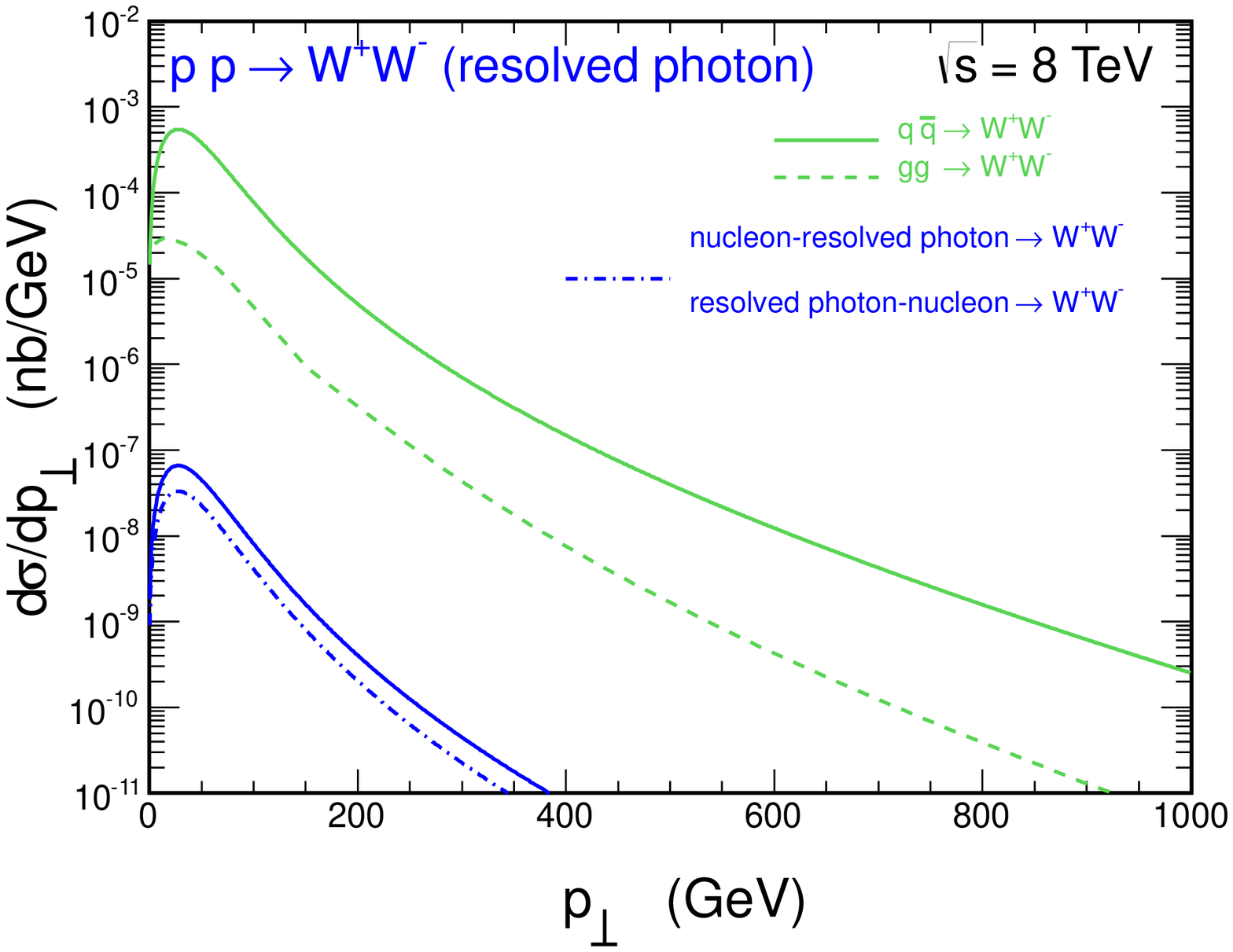}
\includegraphics[width=8cm]{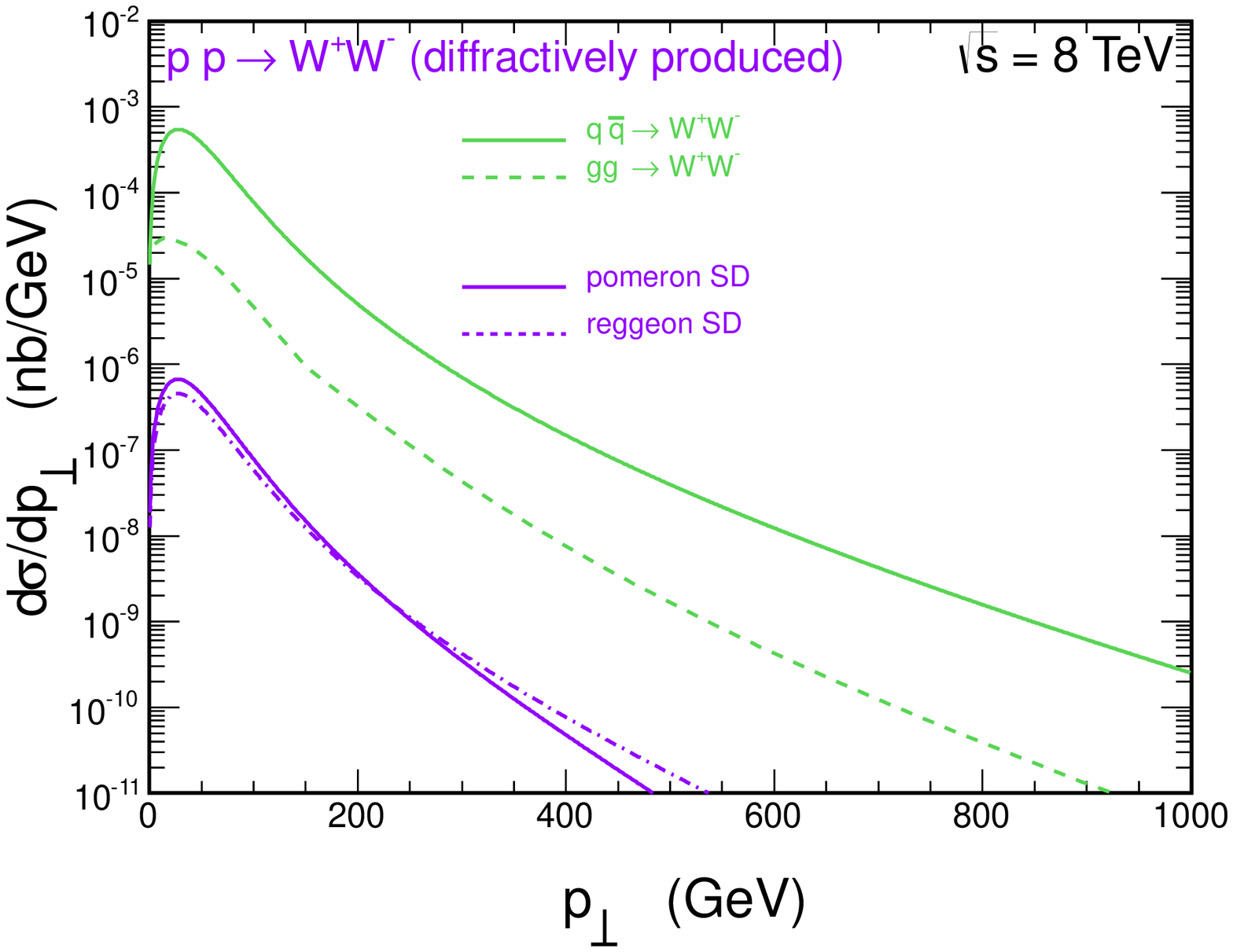}
\end{center}
\caption{ Transverse momentum distribution of $W$ bosons for 
$\sqrt{s}$ = 8 TeV.
The left-top panel shows all photon-photon induced processes, the
right-top panel resolved photon contributions and the bottom panel the
diffractive contribution.
The diffractive cross section has been multiplied by the gap survival
factor $S_G^2 =$ 0.03.
}
\label{fig:dsig_dpt}
\end{figure}

In Fig.\ref{fig:dsig_dpt} we present distribution in transverse momentum
of $W$ bosons. All photon-photon components have rather similar shapes.
The photon-photon contributions are somewhat harder than those for
diffractive and resolved photon mechanisms. 

%

\begin{figure}
\begin{center}
\includegraphics[width=8cm]{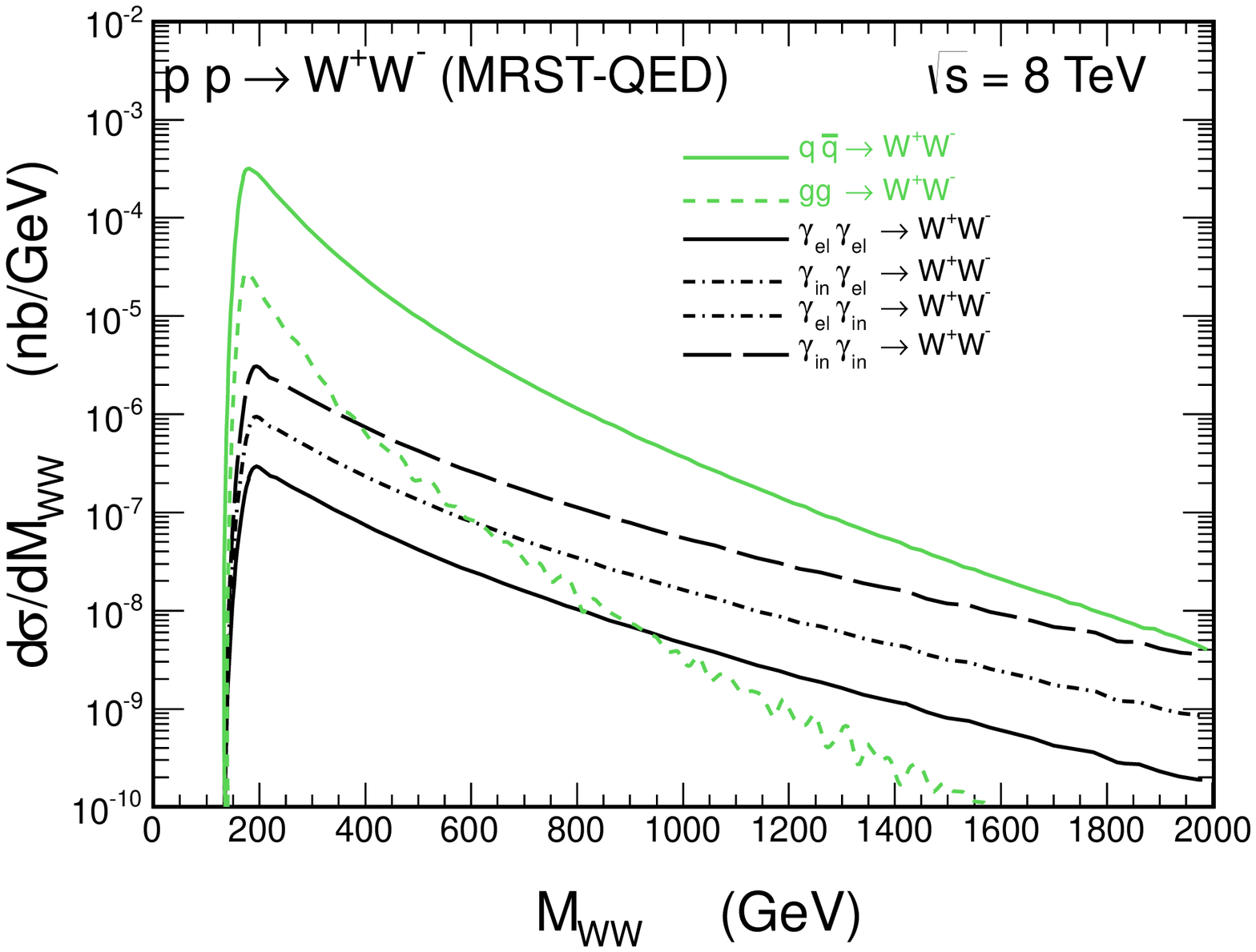}
\includegraphics[width=8cm]{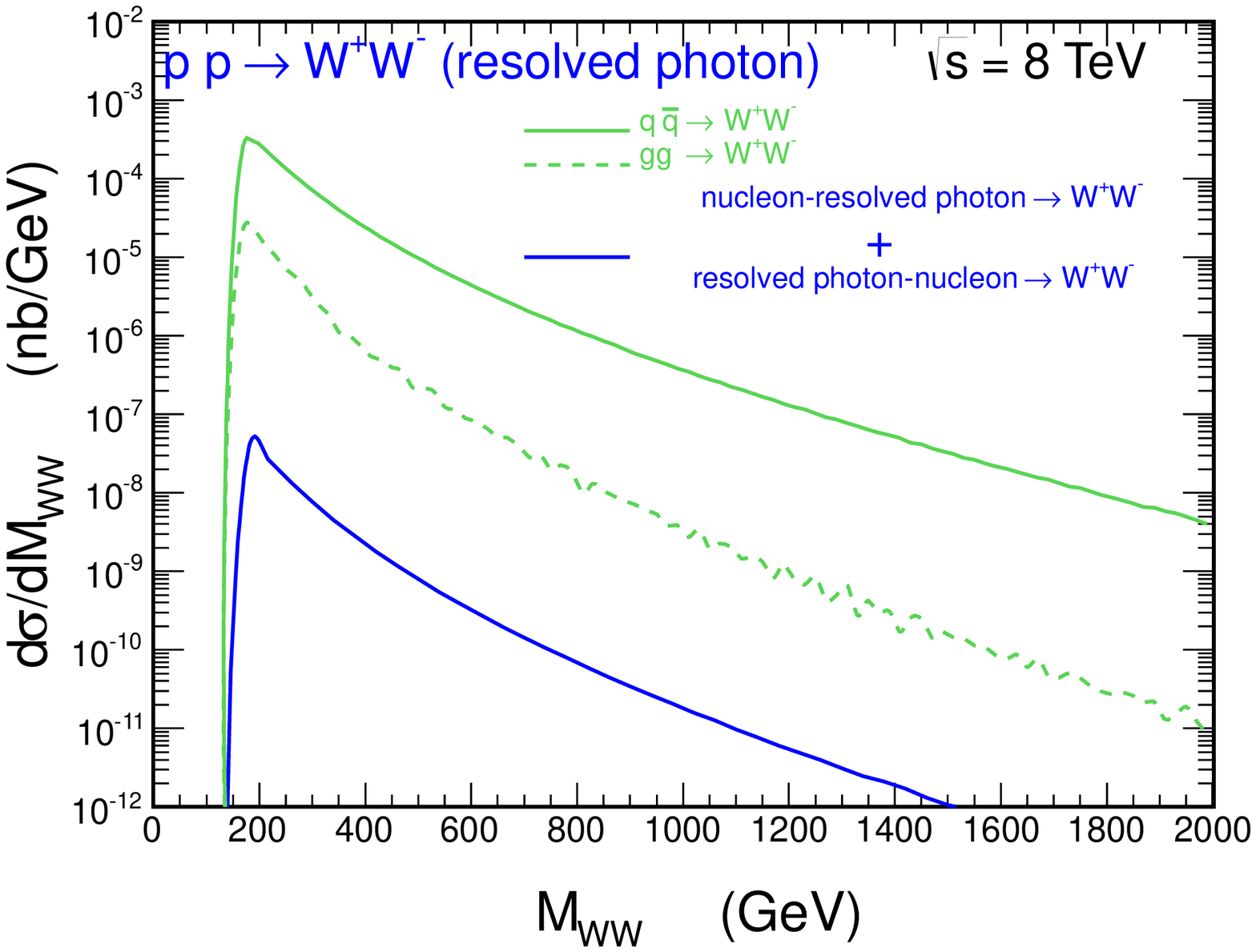}
\includegraphics[width=8cm]{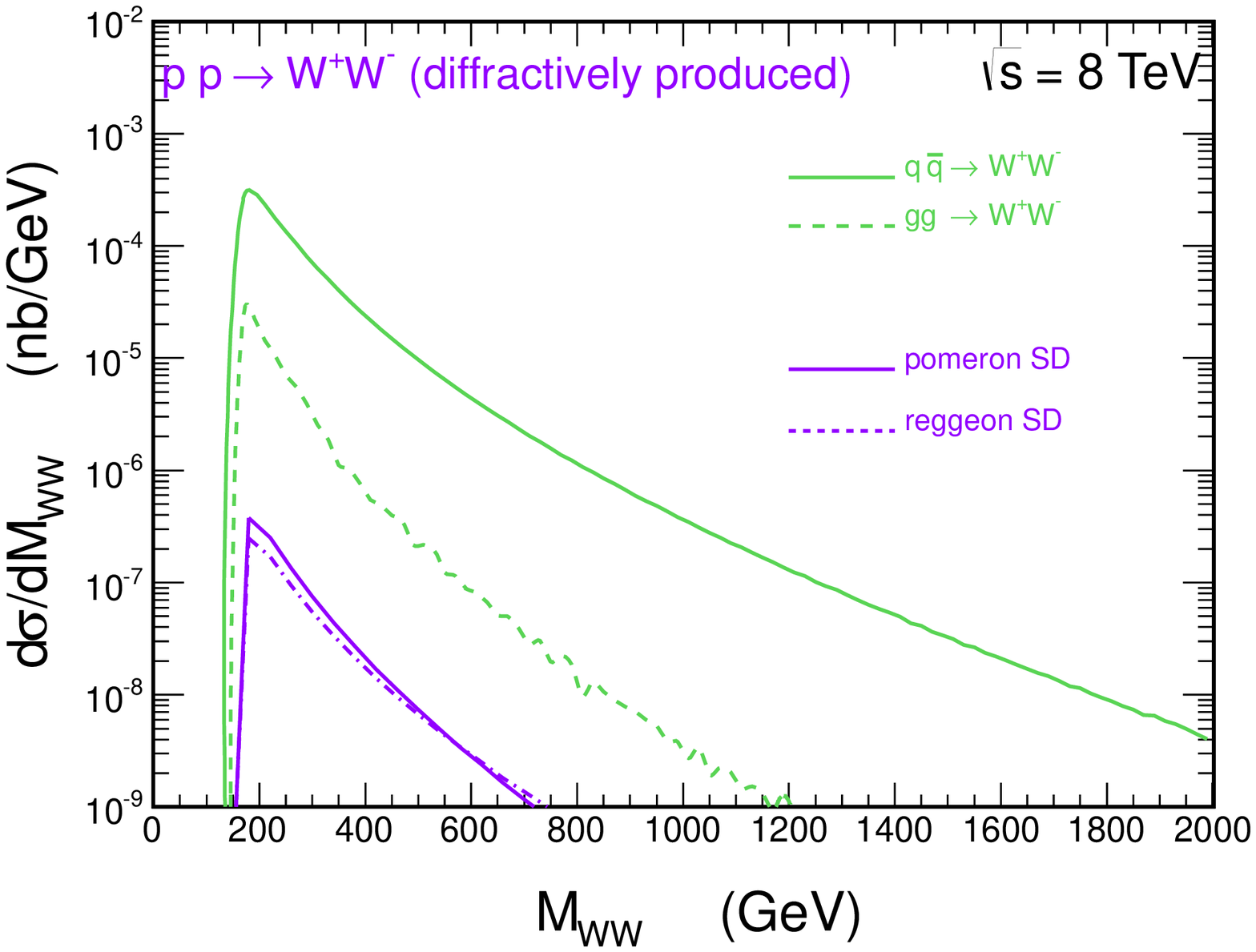}
\includegraphics[width=8cm]{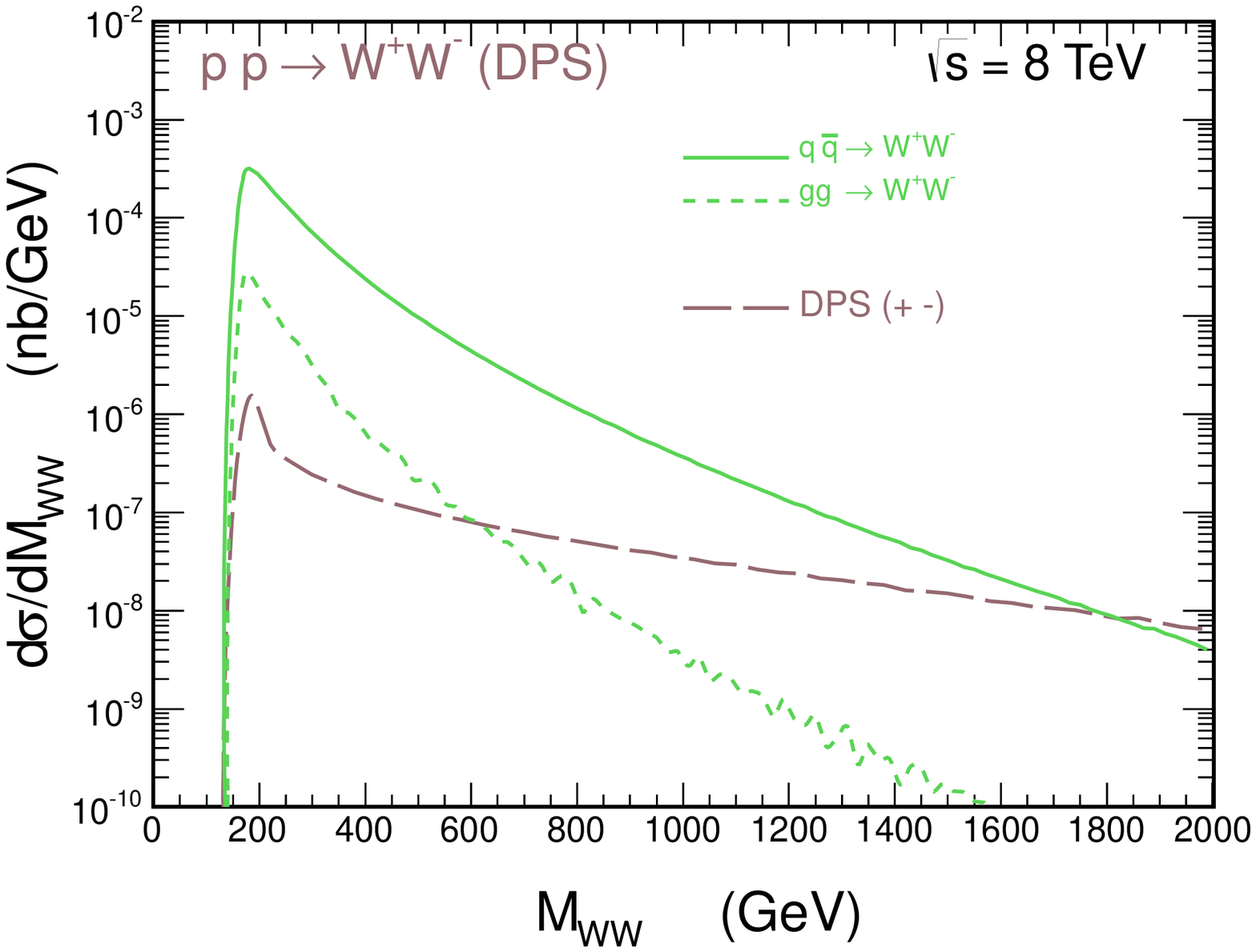}
\end{center}
\caption{ Distribution in invariant mass of the $WW$ pairs for 
$\sqrt{s}$ = 8 TeV.
The top-left panel shows contributions of all photon-photon induced 
processes, the top-right panel resolved photon contributions, 
the bottom-left panel the contributions of diffractive process and 
the bottom-right panel the contribution of DPS.
The diffractive cross section has been multiplied by the gap survival
factor $S_G^2$ = 0.03.
}
\label{fig:dsig_dM}
\end{figure}

In Fig.\ref{fig:dsig_dM} we show distributions in invariant mass of the
$W W$ pairs. The relative contributions of the photon-photon
and DPS components clearly grow with the invariant mass. 
The same would be true for the distribution in the rapidity distance 
between the gauge bosons.
In reality one rather measures charged leptons. Then the distributions
in invariant mass of positive and negative leptons or in rapidity
distance between them would be interesting. This requires a dedicated study
in the future.

\begin{figure}
\begin{center}
\includegraphics[width=5cm]{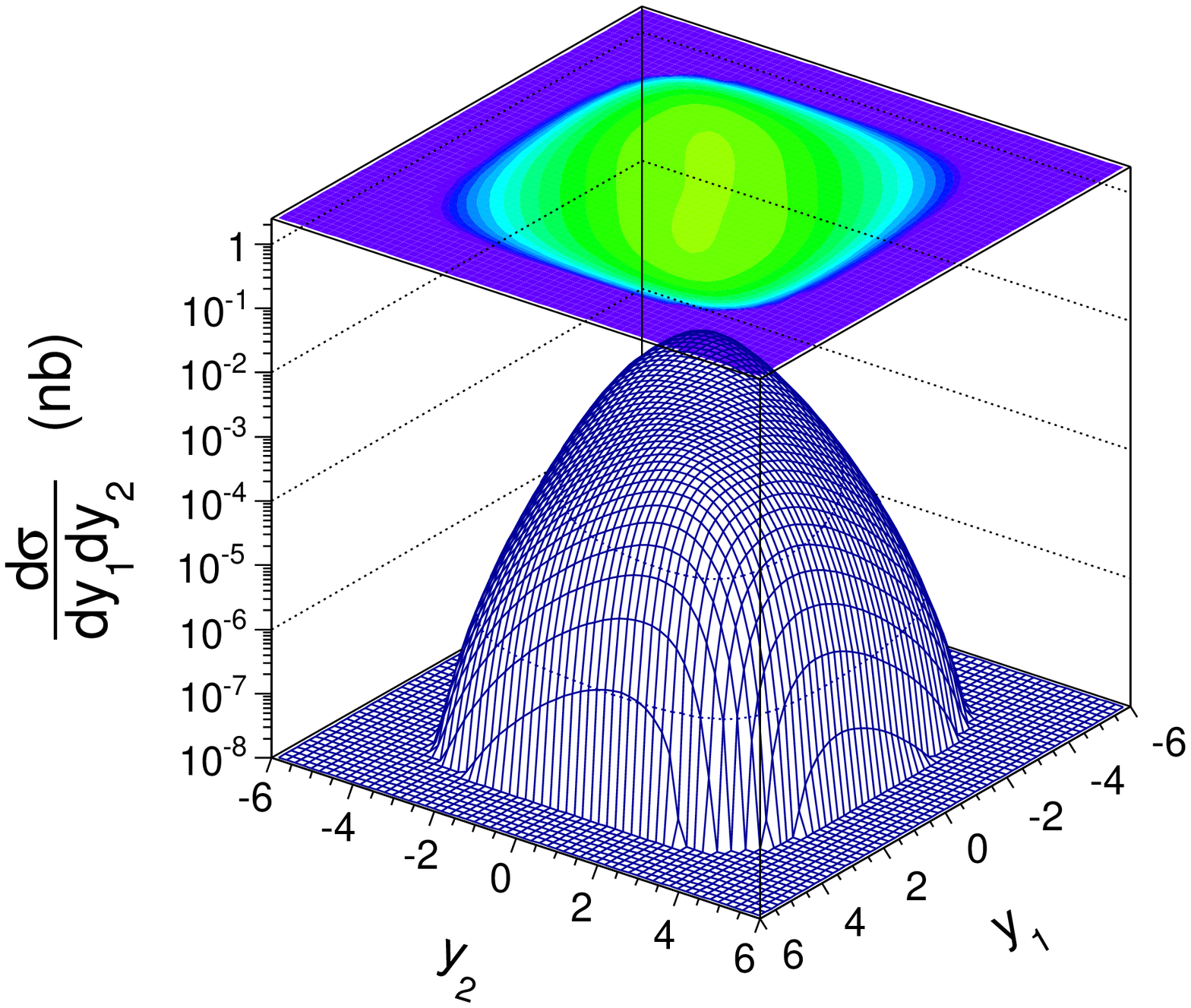}
\includegraphics[width=5cm]{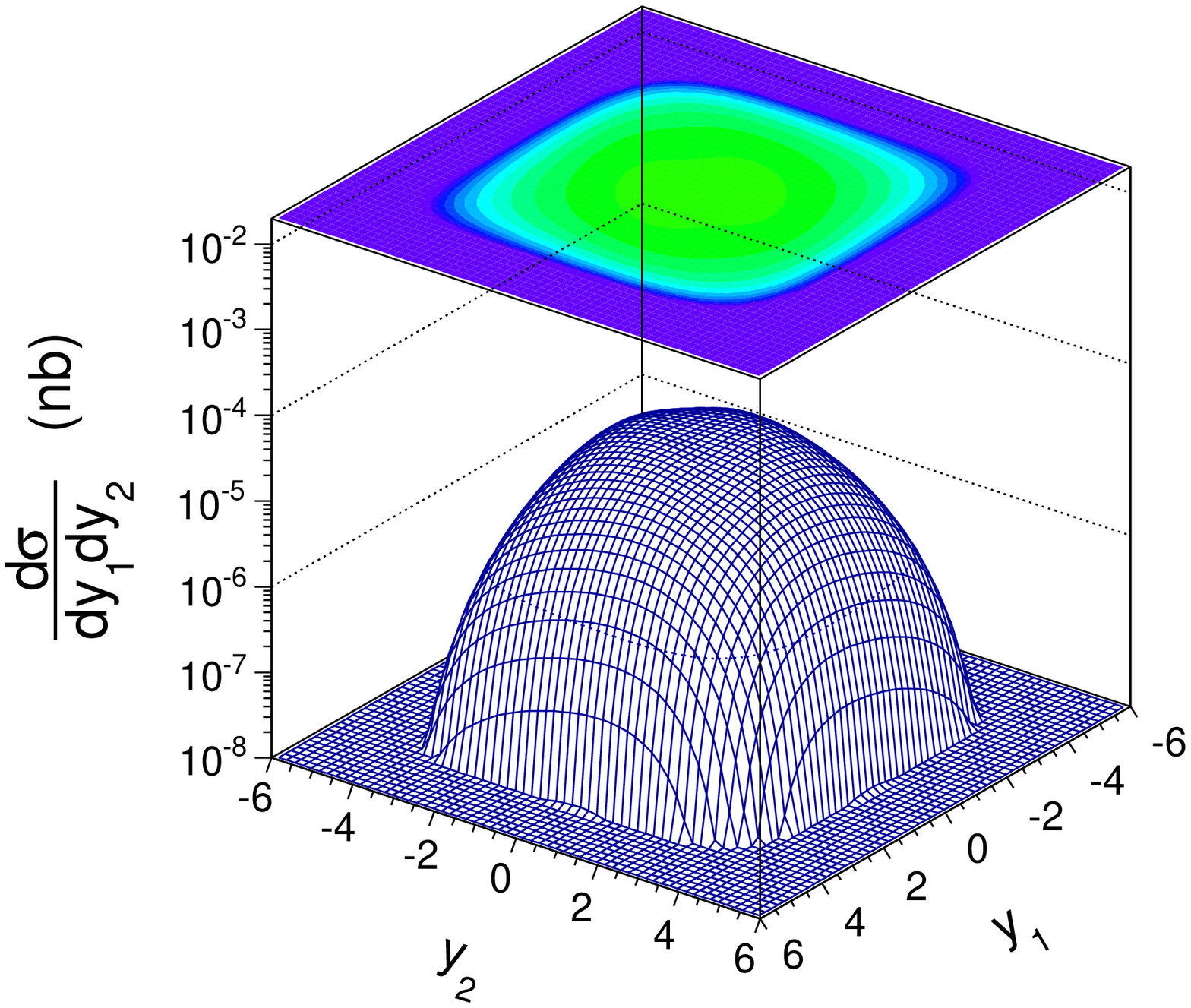}
\includegraphics[width=5cm]{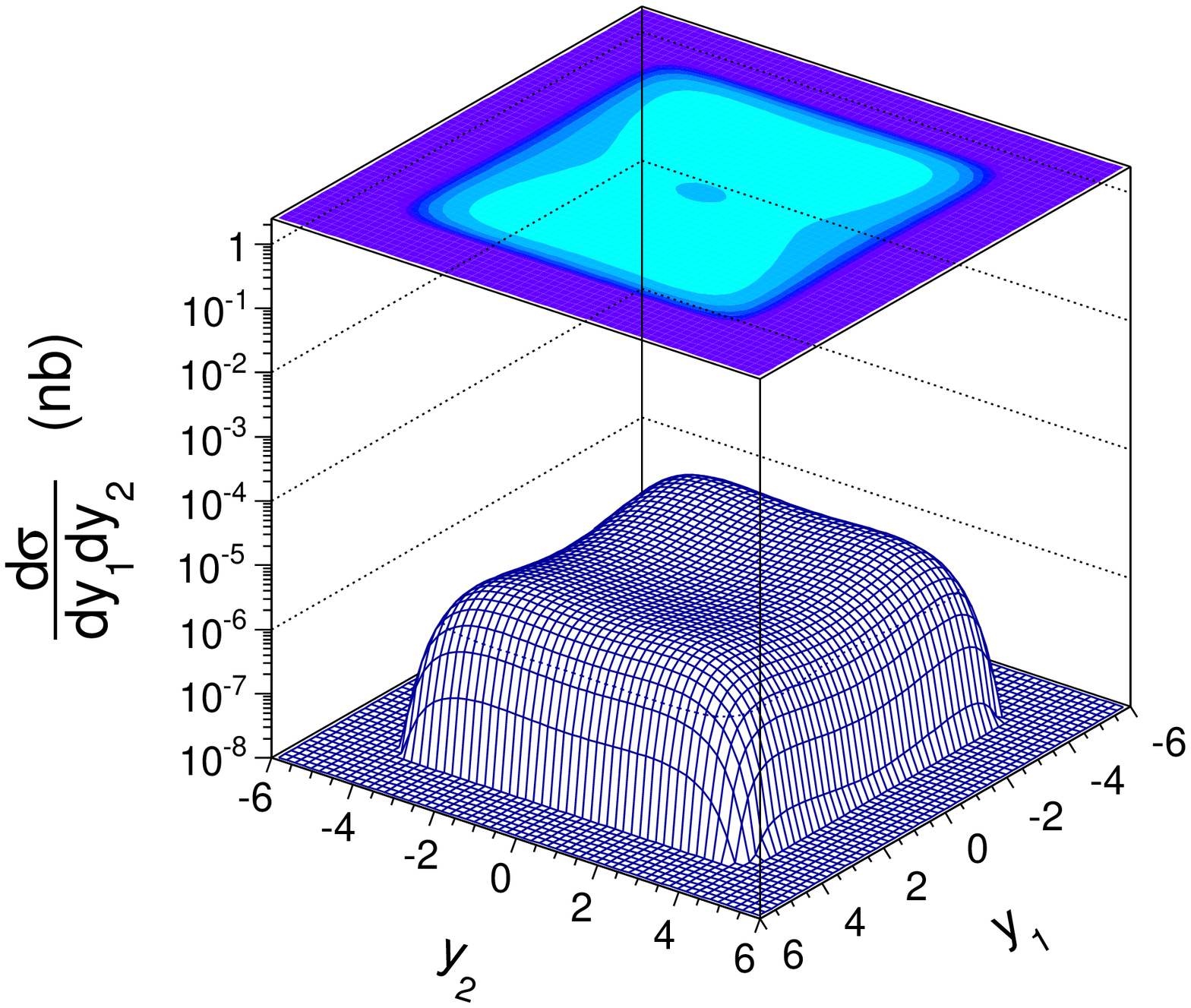}
\end{center}
\caption{ 
Two dimensional distributions in rapidity of $W^+$ and rapidity of $W^-$
for the dominant $q \bar q$ (left), inelastic-inelastic photon-photon
(middle) and double-parton scattering (right) contributions at 
$\sqrt{s}$ = 8 TeV.}
\label{fig:map_y1y2}
\end{figure}

Finally, for completeness, in Fig.\ref{fig:map_y1y2} we present some 
interesting examples of two-dimensional distributions in rapidity of 
$W^+$ and $W^-$. We show three different distributions for the dominant
$q \bar q$, inelastic-inelastic photon-photon and double-parton
scattering components. The $q \bar q$ component dominates
at $y_1, y_2 \approx$ 0. The photon-photon component has ``broader''
distribution in $y_1$ and $y_2$. In contrast, the double-parton
scattering component gives a very flat two-dimensional distribution.
The information presented in the figure can be used in order
to ``enhance'' content of the interesting component. A study of leptons
from the $W$-boson decays would be a next interesting step in
understanding practical possibilities to study the different components
discussed here at the LHC.

The general situation is summarized in Table 1. The photon-photon
induced processes give quite large contribution. The
single-resolved photon contributions are at least order of
magnitude smaller than the diffractive contribution.
This is still surprisingly large. The reason that the single resolved
photon contributions are relatively large is due to the fact that
quark or antiquark carry on average fairly large fraction of the photon
longitudinal momentum. The double parton scattering contribution is
not too large. It may be, however, important for large rapidity
distances between gauge bosons. 
The diffractive contributions in the table are
not multiplied by the gap survival factor ($S^2_G$) which is known 
only approximately.
The single-diffractive contributions
have rather different shape in rapidity than the resolved-photon
contributions in spite of topological similarity.

In real experiments only rather limited part of the phase space is covered.
The total cross section is then obtained by extrapolation into unmeasured
region with the help of Monte Carlo codes.
It is needless to say that none of the new contributions discussed here 
is included when extrapolating purely experimental results to 
the phase-space integrated cross sections. 
This means that the ``measured'' total cross section is underestimated.
In order to answer the question ``how much'' requires dedicated 
Monte Carlo analyses.

\begin{table}
\caption{Contributions of different subleading processes
  discussed in the present paper to the total cross section
for different energies. The cross section is given in pb. The 
diffractive contributions here have not been multiplied by the gap
survival factor. The combinatorial factor 1/2 is shown explicitly for
the double parton scattering (DPS) contribution for $W^+ W^+$ and 
$W^- W^-$ final states shown for completeness.
.}
\begin{tabular}{|c|c|c|c|c|c|}
\hline
contribution               &  1.96 TeV & 7 TeV & 8 TeV & 14 TeV &  comment \\
\hline
CDF                        &  12.1 pb  &       &       &        & \\
D0                         &  13.8 pb &       &       &        & \\
ATLAS                      &       &  54.4 pb  &    &        &  large extrapolation \\
CMS                        &       &  41.1 pb  &    &        &  large extrapolation \\
\hline
$q \bar q$  & 9.86  &  27.24  &   33.04  & 70.21  & dominant (LO, NLO) \\
$g g$     & 5.17 10$^{-2}$  & 1.48  &  1.97  & 5.87 &  subdominant (NLO) \\
\hline        
$\gamma_{el} \gamma_{el}$  & 3.07 10$^{-3}$ & 4.41 10$^{-2}$ & 5.40 10$^{-2}$ & 1.16 10$^{-1}$ &  new, anomalous $\gamma \gamma W W$  \\
$\gamma_{el} \gamma_{in}$  & 1.08 10$^{-2}$ & 1.40 10$^{-1}$ & 1.71 10$^{-1}$ & 3.71 10$^{-1}$ &  new, anomalous $\gamma \gamma W W$  \\
$\gamma_{in} \gamma_{el}$  & 1.08 10$^{-2}$ & 1.40 10$^{-1}$ & 1.71 10$^{-1}$ & 3.71 10$^{-1}$ &  new, anomalous $\gamma \gamma W W$  \\
$\gamma_{in} \gamma_{in}$  & 3.72 10$^{-2}$ & 4.46 10$^{-1}$ & 5.47 10$^{-1}$ & 1.19  &  anomalous $\gamma \gamma W W$ \\
\hline
$\gamma_{el,res} - q/\bar q$ & 1.04 10$^{-4}$  & 2.94 10$^{-3}$  & 3.83 10$^{-3}$ & 1.03 10$^{-2}$ &  new, rather sizeable \\ 
$q/\bar q - \gamma_{el.res}$ & 1.04 10$^{-4}$  & 2.94 10$^{-3}$  & 3.83 10$^{-3}$ & 1.03 10$^{-2}$ &  new, rather sizeable \\    
$\gamma_{in,res} - q/\bar q$ &       &       &       &        &  not calculated \\ 
$q/\bar q - \gamma_{in.res}$ &       &       &       &        &  not calculated \\    
\hline
DPS(++)     &  0.61 10$^{-2}/2$     & 0.22/2              & 0.29/2     &    1.02/2  &  not included in NLO \\
DPS(- -)    &  0.58 10$^{-2}/2$     & 0.76 10$^{-1}/2$    & 0.11/2     &    0.40/2  &  not included in NLO \\
DPS(+-)     &  0.6 10$^{-2}$     & 0.13              & 0.18     &    0.64  &  not included in NLO \\
DPS(-+)     &  0.6 10$^{-2}$     & 0.13              & 0.18     &    0.64  &  not included in NLO \\
\hline
$\Pom p$  $(x_{\Pom} < 0.1)$     &  2.82 10$^{-2}$    &  9.88 10$^{-1}$   &  1.27    &   3.35      &  new, relatively large \\
$p \Pom$  $(x_{\Pom} < 0.1)$     &  2.82 10$^{-2}$    &  9.88 10$^{-1}$   &  1.27    &   3.35      &  new, relatively large \\
$\Reg p$  $(x_{\Reg} < 0.2)$     &  4.51 10$^{-2}$    &  7.12 10$^{-1}$   & 8.92 10$^{-1}$  &  2.22   &  new, relatively large \\
$p \Reg$  $(x_{\Reg} < 0.2)$     &  4.51 10$^{-2}$    &  7.12 10$^{-1}$   & 8.92 10$^{-1}$  &  2.22   &  new, relatively large \\

\hline
\end{tabular}
\end{table}

Some comments on recent studies on $\gamma \gamma W^+ W^-$ boson
couplings, as performed recently by the D0 and CMS collaborations
\cite{gamgam_WW_D0,gamgam_WW_CMS} are in order. In the D0 collaboration 
analysis the inelastic contributions are not included when extracting 
limits on anomalous couplings. 
The CMS collaboration requires an extra condition of no charged
particles in the central pseudorapidity interval.
When comparing calculations to the experimental data the inelastic
contributions are estimated by rescalling elastic-elastic contribution
by an experimental function depending on kinematical variables
(invariant mass, transverse momentum of the $\mu^+ \mu^-$ pair)
obtained in the analysis of $\mu^+ \mu^-$ continuum.
It is not clear to us whether such a procedure is consistent for
$W^+ W^-$ production, where leptons come from decays of the gauge
bosons and invariant mass and transverse moemtum of the $W^+ W^-$ pair
is very different than invariant mass and transverse momentum
of the corresponding dimuons. This cannot be checked in the present 
approach with collinear photons and requires inclusion of photon
transverse momenta.
Within the present approach we predict that inelastic contributions
are significantly larger than the elastic-elastic one. The fragmentation
of the remnants of inelastic excitations is needed to understand to
which extend the inelastic contributions survive the veto condition.
One can expect that particles from the fragmentation of the proton remnants
after the photon emission are emitted in rather forward/backward directions.
The $D0$ collaboration also neglects the inelastic contribution when
calculating the Standard Model background at large lepton transverse 
momenta, in spite they have no explicit veto condition on charge
particles. Inclusion of photon-photon inelastic contributions would
therefore lower considerably their lower limits on parameters of
models with anomalous quartic $\gamma \gamma W W$ coupling. 
The diffractive contributions could also contribute to the distributions 
measured by the CMS and D0 collaborations.
Clearly further analyses that focus on final states of proton remnants 
(after photon emission) are necessary.

\section{Conclusions}

In the present paper we have calculated for the first time a complete set of
photon-photon  and photon-(anti)quark and (anti)quark-photon contributions to
the inclusive production of $W^+ W^-$ pairs. 

The photon-photon contributions can be classified into four topological
categories: elastic-elastic, elastic-inelastic, inelastic-elastic
and inelastic-inelastic, depending whether proton(s) survives 
the emission of the photon or not.
The photon-photon contributions were calculated as done in the past e.g. 
for production of pairs of charged Higgs bosons or pairs of heavy
leptons beyond Standard Model, and within QCD-improved method 
using MRST(QED) parton distributions. 
The second approach was already applied to the production of
Standard Model charged lepton pair production
and $c \bar c$ production.
In the first approach we have obtained: 
$\sigma_{ela,ela} > \sigma_{ela,ine} = \sigma_{ine,ela} >
\sigma_{ine,ine}$.
In the more refined second approach we have got
$\sigma_{ela,ela} <\sigma_{ela,ine} = \sigma_{ine,ela} <
\sigma_{ine,ine}$.
The two approaches give quite different results.
In the first (naive) approach the inelastic-inelastic contribution is
considerably smaller than elastic-inelastic or inelastic-elastic. 
In the approach when photon distribution in the proton undergoes QCD 
$\otimes$ QED evolution, it is the inelastic-inelastic contribution 
which is the biggest out of the four contributions.
This shows that including photon into evolution equation is crucial.
This is also a lesson for other processes known from
the literature, where photon-photon processes are possible.
This includes also some processes beyond the Standard Model mentioned in
this paper.

The inelastic contributions sum up to the cross section of the order 
of 0.5 - 1 pb at the LHC energies. The photon-photon contributions
are particularly important at large $WW$ invariant masses, i.e. 
probably also large invariant masses of charged leptons where 
its contribution is larger than that for gluon-gluon fusion.

The elastic-inelastic or inelastic-elastic contributions are 
interesting by themselves. Since they are related to the emission of 
forward/backward protons they could be potentially measured in the future
with the help of forward proton detectors.
Both CMS and ATLAS have plans for installing such detectors after
the present (2013-2014) shutdown.
Unfortunately the mechanisms are expected to have similar topology of 
the final state as single-diffractive contributions to $W^+ W^-$ production. 
It would be therefore valuable to make a dedicated study how to pin down 
the mixed elastic-inelastic contributions. Clearly this would be a
valuable test of both the formalism presented and our understanding 
of the underlying reaction mechanism.

We have discussed also briefly the double-scattering mechanism
which also significantly contributes to large $M_{WW}$ invariant masses.
This was suggested recently as an important ingredient
for the Higgs background in the $W W^*$ or $Z Z^*$ final channels.
Our estimate is more than order of magnitude smaller than that
suggested recently in the literature in order to explain the Higgs
signal in the $W^+ W^-$ channel.

After this paper was completed we have learned about a detailed study
of uncertainties of a photon PDF in the framework of so-called neural
network PDFs \cite{NNPDF}. This analysis suggests that inelastic
photon-induced contributions may have rather big uncertainies. 
The issue should be better clarified in the future by comparing 
similar calculation for two-photon-induced $\mu^+ \mu^-$ production
at large invariant masses of dileptons to appropriate experimental data.

\vspace{1cm}

{\bf Acknowledgments}

We are indebted to Krzysztof Piotrzkowski, Jonathan Hollar, 
and Gustavo da Silveira for a discussion of the CMS experiment 
on semi-exclusive production of $W^+ W^-$ pairs and 
Christophe Royon, Sudeshna Banerjee and Emilien Chapon for 
the discussion of the D0 studies on $\gamma \gamma W W$ anomalous coupling. 
The help of Piotr Lebiedowicz in calculating the gluon-gluon component
is acknowleded. We are indebted to Wolfgang Sch\"afer for a FORTRAN routine
and an interesting discussion. We are indebted to Stefano Forte and
Juan Rojo for a communication on their recent work on photon NNPDF.
This work was partially supported by the Polish MNiSW grant 
DEC-2011/01/B/ST2/04535 and by the bilateral exchange program
between Polish Academy of Sciences and FNRS, Belgium.




\begin{thebibliography}{100}

\bibitem{royon}
O. Kepka and C. Royon, Phys. Rev. {\bf D78} (2008) 073005;\\
E. Chapon, C. Royon and O. Kepka, Phys. Rev. {\bf D81} (2010) 074003.

\bibitem{piotrzkowski}
N. Schul and K. Piotrzkowski, Nucl. Phys. B (Proc. Suppl.) {\bf 179-180} (2008) 289;\\
T. Pierzcha{\l}a and K. Piotrzkowski, Nucl. Phys. B (Proc. Suppl.) {\bf
  179-180} (2008) 257.

\bibitem{LS2012}
P. Lebiedowicz, R. Pasechnik and A. Szczurek,
Phys. Rev. {\bf D81} (2012) 036003.

\bibitem{WWKhoze}
B.E.~Cox {\it et al.},
Eur. Phys. J. {\bf C45} (2006) 401; \\
S. Heinemeyer {\it et al.},
Eur. Phys. J. {\bf C53} (2008) 231.

\bibitem{Gupta:2011be}
R.S.~Gupta,
Phys.\ Rev.\ {\bf D85} (2012) 014006.

\bibitem{forward_protons}
CMS and TOTEM diffractive and forward physics working group,
CERN/LHCC 2006-039/G-124, CMS Note 2007/002, TOTEM Note 06-5,
December 2006;\\
M. Albrow et al. (FP420 R and D Collaboration), JINST {\bf 4} (2009) T10001
[arXiv:0806.0302 [hep-ex]];\\
The AFP project in ATLAS, Letter of Intent of the PhaseI Upgrade
(ATLAS Collaboration), http://cdsweb.cern.ch/record/1402470; \\
C. Royon (RP220 Collaboration), arXiv:0706.1796 [physics.ins-det];\\
M. Tasevsky, Nucl. Phys. Proc. Suppl. {\bf 179-180} (2008) 187;\\
The PPS upgrade project in CMS, M. Albrow, AIP Conf. Proc.
{\bf 1523} (2012) 320.

\bibitem{CMS2011}
CMS Collaboration, Phys.\ Lett.\ {\bf B699} (2011) 25.

\bibitem{ATLAS2012}
ATLAS Collaboration, arXiv:1203.6232 [hep-ex].

\bibitem{KS2000}
A. Kulesza and W.J. Stirling, Phys. Lett. {\bf B475} (2000) 168.

\bibitem{Kulesza2010}
J.R. Gaunt, Ch.-H. Kom, A. Kulesza and W.J. Stirling,
arXiv:1003.3953 [hep-ph].

\bibitem{GKKS2011}
J.R. Gaunt, Ch.-H. Kom, A. Kulesza and W.J. Stirling,
arXiv:1110.1174 [hep-ph].



\bibitem{KP2013}
W. Krasny and W. P{\l}aczek, arXiV:1305.1769 [hep-ph].

\bibitem{gamgam_WW_CMS}
CMS Collaboration, arXiv:1305.5596 [hep-ex].

\bibitem{gamgam_WW_D0}
D0 Collaboration, arXiv:1305.1258 [hep-ex].

\bibitem{electroweak}
A. Bierweiler, T. Kasprzik, H. K\"uhn and S. Uccirati, JHEP 1211 (2012)
093; \\
A. Bierweiler, T. Kasprzik and J.H. K\"uhn, arXiv:1305.5402;\\
J. Baglio, Le Duc Ninh and M.M. Webber, arXiv:1307.4331.

\bibitem{Higgs_ATLAS}
G.~Aad {\it et al.}  [ATLAS Collaboration],
  Phys.\ Lett.\ {\bf B716}, 1 (2012);
  Science {\bf 338}, 1576 (2012).

\bibitem{Higgs_CMS}
S.~Chatrchyan {\it et al.}  [CMS Collaboration],
  Phys.\ Lett.\ {\bf B716}, 30 (2012);
  Science {\bf 338}, 1569 (2012).

\bibitem{WW_NLO}
J. Ohnemus, Phys. Rev. {\bf D44} (1991) 3477;\\
S. Frixione, P. Nason and G. Ridolfi, Nucl. Phys. {\bf B383} (1992) 3;\\
S. Frixione, Nucl. Phys. {\bf B410} (1993) 280;\\
L.J. Dixon, Z. Kunszt and A. Signer, Nucl. Phys. {\bf B531} (1998) 3;\\
J.M. Campbell, R.K. Ellis, Phys. Rev. {\bf D60} (1999) 113006.

\bibitem{DDS95}
A. Denner, S. Dittmaier and R. Schuster,
Nucl. Phys. {\bf B452} (1995) 80.

\bibitem{DZ}
M. Drees and D. Zeppenfeld,
Phys. Rev. {\bf D39} (1989) 2536.

\bibitem{gg_WW}
D.A. Dicus, C. Kao and W. Repko, Phys. Rev. {\bf D36} (1987) 1570;\\
E.N. Glover and J. van der Bij, Phys. Lett. {\bf B219} (1989) 488;\\
E.N. Glover and J. van der Bij, Nucl. Phys. {\bf B321} (1989) 561.

\bibitem{Eichten}
E. Eichten, L. Hinchliffe, K. Lane, C. Quigg,
Rev. Mod. Phys. {\bf 56} (1984) 579.

\bibitem{DGNR94}
M. Drees, R.M. Godbole, M. Nowakowski and S.D. Rindani,
Phys. Rev. {\bf D94} (1994) 2335.

\bibitem{MRST04}
A.D. Martin, R.G. Roberts, W.J. Stirling, R.S. Thorne,
Eur.Phys.J.{\bf C39}:155-161,2005

\bibitem{IS_ee}
G. Kubasiak and A. Szczurek,
Phys. Rev. {\bf D84} (2011) 014005.

\bibitem{IS_ccbar}
M. {\L}uszczak, R. Maciu{\l}a and A. Szczurek,
Phys. Rev. {\bf D84} (2011) 114018.

\bibitem{H1}
A.~Aktas {\it et al.} [H1 Collaboration],
  Eur.\ Phys.\ J.\  C {\bf 48}, 715 (2006) [arXiv:hep-exp/0606004].

\bibitem{Khoze}
V.A. Khoze, A.D. Martin and M.G. Ryskin, Eur. Phys. J. {\bf C18} (2000)
167.

\bibitem{Maor}
U. Maor, AIP Conf. Proc. {\bf 1105} (2009) 248.

\bibitem{LMS2012}
M. {\L}uszczak, R. Maciu{\l}a and A. Szczurek, 
Phys. Rev. {\bf D85} (2012) 094034.

\bibitem{MS2013}
R. Maciu{\l}a and A. Szczurek, 
Phys. Rev. {\bf D87} (2013) 074039.

\bibitem{NNPDF}
R.D. Ball, V. Bertone, S. Carrazza, L. Del Debbio, S. Forte,
A. Guffanti, N.P. Harltland and J. Rojo,
arXiv:1308.0598.

\end{thebibliography}
\end{document}